\documentclass[twocolumn,apj,floatfix,showkeys]{emulateapj}
\usepackage{latexsym}
\usepackage{epsfig}
\usepackage{amssymb}
\usepackage{amsmath}
\usepackage[latin1]{inputenc}
\usepackage[T1]{fontenc}
\usepackage{amsmath}
\usepackage{pstricks}
\usepackage{graphicx}
\usepackage{hyperref}

\begin{document}


\title{Structure Formation in the Symmetron model}
\author{Anne-Christine Davis$^{2,~\hyperref[mail]{\delta}}$}
\author{Baojiu Li$^{2,3,~\hyperref[mail]{\lambda}}$}
\author{David F. Mota$^{1,~\hyperref[mail]{\mu}}$}
\author{Hans A. Winther$^{1,~\hyperref[mail]{\omega}}$}
\affiliation{$^{1}$Institute of Theoretical Astrophysics,
University of Oslo, 0315 Oslo, Norway} 
\affiliation{$^{2}$DAMTP,
Centre for Mathematical Sciences, University of Cambridge,
Wilberforce Road, Cambridge CB3 0WA, UK}
\affiliation{$^{3}$Kavli
Institute for Cosmology Cambridge, Madingley Road, Cambridge CB3
0HA, UK}

\footnotetext[]{\label{mail}
\\$^{\omega}$	Email address: \href{mailto:h.a.winther@astro.uio.no}{\nolinkurl{h.a.winther@astro.uio.no}}
\\$^{\mu}$		Email address: \href{mailto:d.f.mota@astro.uio.no}{\nolinkurl{d.f.mota@astro.uio.no}}
\\$^{\lambda}$	Email address: \href{mailto:B.Li@damtp.cam.ac.uk}{\nolinkurl{b.li@damtp.cam.ac.uk}}
\\$^{\delta}$	Email address: \href{mailto:A.C.Davis@damtp.cam.ac.uk}{\nolinkurl{a.c.davis@damtp.cam.ac.uk}}}


\begin{abstract}
Scalar fields, strongly coupled to matter, can be present in
nature and still be invisible to local experiments if they are
subject to a screening mechanism. The symmetron is one such
mechanism which relies on restoration of a spontaneously broken
symmetry in regions of high density to shield the scalar
fifth force. We have investigated structure formation in the
symmetron model by using N-body simulations and find strong
observable signatures in both the linear and nonlinear matter
power spectrum and on the halo mass function. The mechanism for
suppressing the scalar fifth force in high density regions is also
found to work very well.
\end{abstract}


\maketitle


\section{Introduction}

Our current standard model of cosmology, $\Lambda$CDM, has been
very successful in explaining a large range of observations
probing a vast range in length scales. We should nevertheless be
open for the possibility that $\Lambda$CDM is just a first order
approximation of some more fundamental theory. Many theories of
high energy physics, like string theory and supergravity, predict
light gravitationally coupled scalar fields (see e.g.
\cite{supersymmetry_review,linde} and references therein). These
scalars may play the role of dark energy (quintessence). If these
scalar fields have non-minimal coupling to matter fields, then
they could mediate extra forces which are potentially detectable
in local experiments.

Over the last decades, several laboratory and solar system
experiments have tried to detect a sign of such fundamental
coupled scalar fields \citep{eotwash,hoskins,decca,cassini}, but
the results so far show no signature of them. Naively, the results
of these experiments have ruled out any such scalar fields.
However, one should bear in mind that a coupled scalar field might
exist but is undetected just because it is either very weakly
coupled or very heavy.

To this day we know three types of theoretical mechanisms (see
\cite{khoury_screening_mechanisms} for a review) that can explain
why such light scalars, if they exist, may not be visible to
experiments performed near the Earth. One such class, the
chameleon mechanism
\citep{chameleon_cosmology,cosmological_chameleon,alpha3,alpha1,alpha2}, operates when
the scalars are coupled to matter in such a way that their
effective mass depends on the local matter density. In space,
where the local mass density is low, the scalars would be light
and deviations from General Relativity would be observed. But near
the Earth, where experiments are performed, the local mass density
is high and the scalar field would acquire a heavy mass making the
interactions short range and therefore unobservable.

The second mechanism, the Vainshtein mechanism
\citep{Vainshtein,dvali_gabadadze_vains,schwartz_massivegravity},
operates when the scalar has derivative self-couplings which
become important near matter sources such as the Earth. The strong
coupling near sources essentially cranks up the kinetic terms,
which translates into a weakened matter coupling. Thus the scalar
screens itself and becomes invisible to experiments. This
mechanism is central to the phenomenological viability of
braneworld modifications of gravity and galileon scalar theories
\citep{dgp,cascading_dgp,galileon_modifiedgravity,multifield_galileons,gal1,gr_aux_dim,massivegravity,davis_gal}.

The last mechanism, the one explored in this paper, is the
symmetron mechanism
\citep{khoury_symmetron,symmetron_cosmology,olive_pospelov,symmetron_brax}. In
this mechanism, the vacuum expectation value (VEV) of the scalar
depends on the local mass density, becoming large in regions of
low mass density, and small in regions of high mass density. By
taking the coupling of the scalar to matter to be proportional to
the VEV, we can have a viable theory where the scalar couples with
gravitational strength in regions of low density, but is decoupled
and screened in regions of high density. This is achieved through
the interplay of a symmetry breaking potential and a universal
quadratic coupling to matter.

In vacuum, the scalar acquires a VEV which spontaneously breaks
the $\mathbb{Z}_2$ symmetry $\phi\to-\phi$. In the regions of
sufficiently high matter density, the field is confined near $\phi
= 0$, and the symmetry is restored. The fifth force arising from
the matter coupling is proportional to $\phi$ making the effects
of the scalar small in high density regions.

As opposed to Chameleons, where the strongest constraints
\citep{mota_shaw,brax_generalchameleon,brax_shaw_fofr,chameleon_signatures,cham1,cham2,cham3,cham4}
comes from laboratory experiments which in effect washes out any
observable effects in the solar system, the symmetron predicts a
host of observational signatures in experiments designed to look
for deviations from GR, which are just below the currents bounds
and within reach of the next generation experiments.

The cosmology of coupled scalar field models are usually strongly
constrained by local gravity experiments, which could put limits
on the range and the coupling strength of the scalar field. There
do exist several cases in which signatures on the linear
perturbations are found, but in most cases the range of the field
is well below linear scales. To proceed into the region of
nonlinear structure formation one can use the spherical collapse
model to obtain the qualitative behavior, but in order to obtain
accurate quantitative results deep into the nonlinear regime one
is almost required to perform N-body simulations.

Studies of coupled scalar field models by using N-body simulations
\citep{baojiu_fofr,dilaton,baojiu_coupled_sfc,baojiu_scalar_structure_effects,baojiu_extended_quintessence,baojiu_powerspectrum_nbody,baojiu_differentcouplings,baldi_1,baldi_2}
have revealed several interesting signatures which can in
principle be detected by observations in the near future. For example, in
\cite{baojiu_enviroment_halos} it was found that $f(R)$ theories
can give rise to a dependence on the environment of the
dynamical to lensing mass ratio of halos; an observable feature
that is not found in $\Lambda$CDM.

In this article we will study the effects a symmetron field has on
structure formation. By performing high resolution N-body
simulations we demonstrate explicitly how the Symmetron mechanism
works in screening the fifth force and obtain observables as the
matter power spectrum and the mass function.


\section{The Symmetron Model}

The action governing the dynamics of the symmetron model is given by
\begin{align}\label{action}
S =& \int dx^4 \sqrt{-g}\left[\frac{R}{2}M_{\rm pl}^2 - \frac{1}{2}(\partial\phi)^2 -
V(\phi)\right]\nonumber\\
&+ S_m(\tilde{g}_{\mu\nu},\psi_i)
\end{align}
where $g$ is the determinant of the metric $g_{\mu\nu}$, $R$ is
the Ricci scalar, $\psi_i$ are the different matter fields and
$M_{\rm pl} \equiv \frac{1}{\sqrt{8\pi G}}$ where $G$ is the bare
gravitational constant. The matter fields couple to the
Jordan frame metric $\tilde{g}_{\mu\nu}$ via a conformal rescaling
of the Einstein frame metric $g_{\mu\nu}$ given by
\begin{equation}\label{conformal_coupling}
\tilde{g}_{\mu\nu} = A^2(\phi)g_{\mu\nu}
\end{equation}
The coupling function $A(\phi)$ is chosen to be an even polynomial in $\phi$ (to be compatible with the $\phi\to-\phi$ symmetry)
\begin{equation}\label{coupling_function}
A(\phi) = 1 + \frac{1}{2}\left(\frac{\phi}{M}\right)^2 + \mathcal{O}\left(\frac{\phi^4}{M^4}\right)
\end{equation}
described by a single mass scale $M$. For the range of parameters we are interested in we have $\left(\frac{\phi}{M}\right)^2\ll 1$, thus, we can neglect the higher order correction terms. The potential is chosen to be of the symmetry breaking form
\begin{equation}\label{potential}
V(\phi) = V_0 -\frac{1}{2}\mu^2\phi^2 + \frac{1}{4}\lambda\phi^4
\end{equation}
where $V_0$ is a cosmological constant (CC). We will for
simplicity absorb all contributions to the CC into $V_0$ by simply
putting $V_0 \equiv \Lambda$. We will later see that $\Lambda$
must be taken to be the usual CC to obtain late time acceleration
of the Universe. The field equation for $\phi$ follows from the
variation of the action Eq.~(\ref{action}) with respect to $\phi$
and reads
\begin{equation}\label{eom_phi}
\square\phi = V_{\rm eff,\phi}
\end{equation}
The effective potential is given in terms of the trace, $T_m$, of
the matter energy-momentum tensor by
\begin{align}\label{veff}
V_{\rm eff}(\phi) &=  \frac{1}{2}\left(-\frac{T_m}{M^2}-\mu^2\right)\phi^2 + \frac{1}{4}\lambda\phi^4\\
&= \frac{1}{2}\left(\frac{\rho_m}{M^2}-\mu^2\right)\phi^2 + \frac{1}{4}\lambda\phi^4
\end{align}
depending on the two mass scales $\mu$, $M$ and the dimensionless
coupling constant $\lambda$. It is convenient to define the
critical matter density (and the critical redshift)
\begin{align}\label{rho_crit}
\rho_{\rm SSB} &\equiv \mu^2 M^2 = 3H_0^2M_{\rm pl}^2\Omega_m(1+z_{\rm SSB})^3
\end{align}
where $\Omega_m$ is the matter density parameter in the Universe
today and $H_0$ is the Hubble parameter. In regions where $\rho_m
> \rho_{\rm SSB}$ (where $\rho_m$ is the local matter density) the symmetry $\phi\to-\phi$ is upheld and the
effective potential has a minimum at $\phi_{\rm min} = 0$, whereas
in regions where $\rho_m < \rho_{\rm SSB}$ the symmetry is
spontaneously broken and the field acquires a VEV
\begin{equation}\label{phi0}
\phi_{\rm min} = \pm\phi_0 \sqrt{1-\frac{\rho_m}{\rho_{\rm SSB}}}
\end{equation}
where $\phi_0 \equiv \frac{\mu}{\sqrt{\lambda}}$ is the symmetry breaking VEV for $\rho_m\to 0$. The mass of small fluctuations around the minimum of the effective potential is given by
\begin{align}\label{mass}
m_{\phi}^2 &\equiv V_{\rm eff,\phi\phi} = \left(\frac{\rho_m}{\rho_{\rm SSB}}-1\right)\mu^2 + 3\lambda\phi_{\rm min}^2\nonumber\\
& = \left\{\begin{array}{cc} \mu^2\left(\frac{\rho_m}{\rho_{\rm SSB}}-1\right), & \rho_m > \rho_{\rm SSB}\\2\mu^2\left(1-\frac{\rho_m}{\rho_{\rm SSB}}\right),& \rho_m < \rho_{\rm SSB}\end{array}\right.
\end{align}
The symmetron field acquires the longest range, $\lambda_{\phi}
\equiv \frac{1}{m_{\phi}}$, in low density regions where
\begin{align}\label{range}
\lambda_{\phi} = \lambda_0 \equiv \frac{1}{\sqrt{2}\mu}
\end{align}
For future convenience we introduce the dimensionless quantity $L
\equiv \frac{\lambda_0}{\text{Mpc}/h}$, which is the maximum range
of the symmetron mediated force in units of Mpc$/h$.
\\\\
The gravitational field equation for $g_{\mu\nu}$ is given by
\begin{align}\label{eeq}
G_{\mu\nu} = 8\pi GT_{\mu\nu}
\end{align}
where the total energy-momentum tensor $T_{\mu\nu}$ is the sum of
the matter and scalar field parts:
\begin{align}\label{ttot}
T_{\mu\nu} = A(\phi)T^m_{\mu\nu} + \phi_{;\mu}\phi_{;\nu} - g_{\mu\nu}\left(\frac{1}{2}(\partial\phi)^2 + V(\phi) \right)
\end{align}
Note that the matter part itself is not conserved, but instead
satisfies
\begin{align}
\nabla_{\nu}T_m^{\mu\nu} = \frac{d\log A(\phi)}{d\phi}\left(T_m\nabla
^{\mu}\phi - T_m^{\mu\nu}\nabla_{\nu}\phi\right)
\end{align}
In N-body simulations we are interested in describing the matter sector by particles and the energy-momentum tensor of an individual particle with mass $m_0$ at position $\bf{r}_0$ is given by
\begin{align}
T_m^{\mu\nu}({\bf r}) = \frac{m_0}{\sqrt{-g}}\delta({\bf r}-{\bf r}_0)\dot{r}_0^{\mu}\dot{r}_0^{\nu}
\end{align}
where ${\bf r}$ is the general spatial coordinate. Taking the divergence of Eq.~(\ref{eeq}) and using the Bianchi identity we get the geodesic equation for the matter particles
\begin{align}\label{geo}
\ddot{r}_0^{\mu} + \Gamma^{\mu}_{\alpha\gamma}r_0^{\alpha}r_0^{\gamma} = -\frac{d\log A(\phi)}{d\phi}\left(\nabla^{\mu}\phi + \dot{\phi}\dot{r}_0^{\mu}\right)
\end{align}
which for $A \equiv 1$ reduces to the standard geodesic equation in general relativity.


\subsection{The Symmetron Mechanism: Local Constraints}\label{local_constraints}

From Eq.~(\ref{geo}) we see that the symmetron field gives rise to
a fifth force on the matter fields which, in the nonrelativistic
limit, is given by
\begin{align}\label{fifth force}
\vec{F}_{\phi} = \frac{\phi}{M^2}\vec{\nabla}\phi = \frac{\beta}{M_{\rm pl}} \left(\frac{\phi}{\phi_0}\right)\vec{\nabla}\phi
\end{align}
where we have introduced the coupling constant $\beta \equiv
\frac{\phi_0 M_{\rm pl}}{M^2}$.

The static spherical symmetric solutions of the field equations
were found in \cite{khoury_symmetron}. For two test masses in a region where $\phi = \phi_B$ it was shown that
that the fifth force is simply
\begin{align}\label{fifth force1}
\frac{F_{\phi}}{F_N} = 2\beta^2\left(\frac{\phi_B}{\phi_0}\right)^2
\end{align}
In a low density region ($\rho \ll \rho_{\rm SSB}$) we have $\phi_B = \phi_0$ and the fifth force is comparable
with gravity for  $\beta = \mathcal{O}(1)$.

For very large bodies, the situation is
quite different. The symmetry is restored in the interior of the
body and the fifth force on a test mass outside becomes
\begin{align}\label{fifth_force2}
\frac{F_{\phi}}{F_N} = 2\beta^2\left(\frac{\phi_B}{\phi_0}\right)^2 \frac{1}{\alpha},~~~~\alpha^{-1} = 2\frac{\rho_{\rm SSB}}{\rho_{\rm body}}\left(\frac{\lambda_0}{R_{\rm body}}\right)^2
\end{align}
The fifth force is suppressed by a factor $\alpha^{-1} \ll
1$ $-$ similar to the thin shell factor found in chameleon theories \citep{chameleon_cosmology}.

We also see that if the test masses are inside a screened region $\left(\frac{\phi_B}{\phi_0} \ll 1\right)$ the force will be further suppressed.

Since the field is long ranged (and universally coupled) in
almost all situations today the theory is
best constrained by solar system experiments which have been
performed with high precision.

It turns out that as long as our Galaxy is sufficiently screened ($10\lesssim \alpha_G$), our Sun will also be screened and the combined effects discussed above are enough to evade the current parameterized post Newtonian (PPN) constraints. 

By assuming that $\phi\to\phi_0$ outside
our Galaxy, i.e. that our galactic neighborhood is not screened, these constraints were derived in
\cite{khoury_symmetron,symmetron_brax}, and require
\begin{align}\label{exp_const}
 M \lesssim 10^{-3}M_{\rm pl}
 \end{align}
If the assumption about the value of $\phi$ outside our Galaxy,
which is very likely to be true, can be relaxed then the bound
above can be relaxed somewhat as well. The constraint on $M$ turns
into a constraint on the range of the field and the redshift in
which the SSB takes place:
\begin{align}\label{constraint}
\lambda_0 \lesssim 2.3 \sqrt{\frac{0.3}{\Omega_m}}(1+z_{\rm SSB})^{-3/2} \text{Mpc}/h
\end{align}
Thus for transitions that take place close to the present, the
fifth force can have a range of at most a few Mpc$/h$.


\subsection{Physical Parameters}

In the rest of this article, instead of working with the
parameters $\{\mu,M,\lambda\}$, we will instead choose to work
with the more physically intuitive quantities $\{L,\beta, z_{\rm
SSB}\}$ : the cosmological range of the fifth force in Mpc$/h$,
the strength of the fifth force relative to gravity and the
redshift at which the SSB takes place in the cosmological
background.

The transformation between the two sets of parameters is given by
\begin{align}\label{parameter_conversion}
\frac{\mu}{H_0} &= \frac{2998}{\sqrt{2} L}\\
\frac{M}{M_{\rm pl}} & = 10^{-3}\sqrt{\frac{\Omega_m}{0.27}}\left(\frac{L}{2.36}\right)(1+z_{\rm SSB})^{3/2}\\
\lambda &= \left(\frac{10^{60}H_0}{M_{\rm pl}}\cdot\frac{0.27}{\Omega_m}\right)^2 \frac{1.38\cdot10^{-100}}{\beta^2 L^6(1+z_{\rm SSB})^{6}}
\end{align}
For typical parameters $L\sim \beta \sim 1$ and $z_{\rm SSB} \sim
0$ we have $\mu \sim 10^{-57} M_{\rm pl}$, $M\sim 10^{-3}M_{\rm
pl}$ and $\lambda \sim 10^{-100}$: thus the symmetron is very
weakly coupled.

We will choose to work with values of the parameters, that are
close to the local constraints, and in which the symmetron can
produce observable cosmological effects. This means we will be
most interested in the parameter space $L = \mathcal{O}(1)$,
$\beta = \mathcal{O}(1)$ and $0 \lesssim z_{\rm SSB} \lesssim 2$.


\section{Symmetron Cosmology}

In this section we discuss the cosmological evolution of the
symmetron field from the background evolution to linear
perturbations and derive the nonrelativistic limits of the field
equations to be implemented in the N-body code. The analysis in this section is mainly for comparison with the N-body simulations. For a more thorough discussion regarding the background cosmology and linear perturbations in the symmetron see \cite{symmetron_cosmology} and \cite{symmetron_brax} respectively.

\subsection{Background Cosmology}

The background evolution of the symmetron in a flat
Friedmann-Lemaitre-Robertson-Walker (FRLW) metric
\begin{align}
ds^2 = -dt^2 + a^2(t)(dx^2+dy^2+dz^2)
\end{align}
is determined by the field equation
\begin{align}
\ddot{\phi} + 3H\dot{\phi} + V_{\rm eff, \phi} = 0
\end{align}
together with the Friedman equations
\begin{align}
&3H^2M_{\rm pl}^2 = \rho_m A(\phi) + \rho_{\phi}\\
&\dot{\rho}_m + 3H\rho_m = 0
\end{align}
where
\begin{align}
\rho_{\phi} = \Lambda-\frac{1}{2}\mu^2\phi^2+\lambda\phi^4+ \frac{1}{2}\dot{\phi}^2
\end{align}
When the field follows the minimum of the effective potential we
have
\begin{align}
\left|\frac{\rho_{\phi}-\Lambda}{\Lambda}\right| &\lesssim \frac{\mu^4}{\lambda \Lambda} = \beta^2\frac{\rho_{\rm SSB}}{\Lambda}\left(\frac{M}{M_{\rm pl}}\right)^2\\
&\lesssim 10^{-6}\beta^2(1+z_{\rm SSB})^3
\end{align}
thus for $\beta,z_{\rm SSB} \sim \mathcal{O}(1)$ the dynamical
part of the potential is too small to contribute significantly to
the energy density of the Universe and we are left with the
cosmological constant to account for dark energy.

In the same regime, the coupling function $A(\phi)$ satisfies
\begin{align}
|A(\phi)-1| = \frac{1}{2}\left(\frac{\phi}{M}\right)^2 \lesssim \beta^2\left(\frac{M}{M_{\rm pl}}\right)^2 \lesssim 10^{-6}\beta^2
\end{align}
which is also too small to produce an observable effect on the
background expansion. This implies that the symmetron evades Big
Bang Nucleosyntesis (BBN) bounds on the variation of masses of the standard
model particles (see Sec.~(\ref{BBN_bounds})). It might be
possible to make the symmetron responsible for dark energy by
changing the form of the potential and coupling. One such
modification was proposed in \cite{symmetron_cosmology}, however
it was shown that tuning of the parameters was required to yield
the desired late time cosmology.

\begin{figure}
\centering
\centerline{\includegraphics[width=\columnwidth]{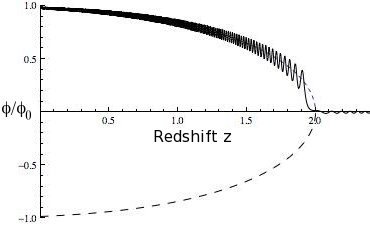}}
\caption{The background evolution of the symmetron for $\beta = 1$, $L=1$ and $z_{\rm SSB} = 2$ together with the analytical background (dashed lines). The symmetry is broken at $z = 2$ and the field settles at one of the two branches.}
\label{phi_back}
\end{figure}

In Fig.~(\ref{phi_back}) we see the background evolution $\phi(z)$
for $z_{\rm SSB} = 2$ together with the analytical minimum. Notice
that the field does not immediately start to follow the minimum
right after SSB. This has important consequences for the evolution
of the perturbations which will be discussed in section
\ref{tac_ins}.

\subsection{Linear perturbations}\label{linear_pert_section}

The most general metric in a perturbed FLRW space-time is given by
\begin{align}
ds^2=-(1+2\alpha){dt}^2-2 a B_{,i}{dt}{dx}^i\\
+a^2\left((1+2\psi)\delta_{ij}+2\gamma_{,i;j}\right){d x}^i{d x}^j\nonumber
\end{align}
where the covariant derivative is given in terms of the
spatial metric, which in the case of our flat background
reduces to $\delta_{ij}$. We decompose the field $\phi$ into the
background and perturbated parts: $\phi({\bf x},t) =
\overline{\phi}(t) +\delta\phi({\bf x},t)$. The energy-momenntum
tensor of nonrelativistic matter can be decomposed as
\begin{align}
T^0_0 = -\rho_m(1+\delta_m), ~~~T^0_i = -\rho_m v,_i
\end{align}
where $v$ is the peculiar velocity of nonrelativistic matter and
$\delta_m$ is the matter density perturbation defined by
\begin{equation}
\delta_m\equiv \frac{\delta \rho_m}{\rho_m}-\frac{\dot{\rho}_m}{\rho_m}v\equiv \frac{\delta\rho_m}{\rho_m}~~~\text{in the comoving gauge}
\end{equation}
The equation determining the evolution of the perturbations, neglecting anisotropic stresses, follows from the Einstein equations. The scalar perturbations can be read off from the formulation of \citep{hwang}, which is independent of 
gauge. In the following we use units of $M_{\rm pl} \equiv 1$. After solving for the different metric potentials we find that the scalar perturbations, in the comoving gauge ($v=0$) are determined by

\begin{align}\label{deltam_eq}
&\ddot{\delta}_m+2H\dot{\delta}_m-\frac{1}{2}\rho_m\delta_m\\
&-\frac{\phi\delta\phi}{M^2}\left(6H^2+6\dot{H}+\Omega_mH^2-\frac{k^2}{a^2}+2\dot{\phi}^2\right)\nonumber\\
&-\frac{\phi}{M^2}\left(\ddot{\delta\phi}+5H\dot{\delta\phi}\right) - \frac{2\dot{\phi}}{M^2}\left(\dot{\delta\phi}+H\delta\phi\right)\nonumber\\
& + V_{\rm eff,\phi}\left(1+\frac{1}{M^2}\right)\delta{\phi}-2\dot{\phi}\dot{\delta\phi}=0\nonumber
\end{align}
\begin{align}\label{phi_eq}
&\ddot{\delta\phi}+\left(3H+\frac{2\phi\dot{\phi}}{M^2}\right)\dot{\delta \phi}+\frac{\phi\rho_m\delta_m}{M^2}-\dot{\phi}\dot{\delta}_m\\
&+\left(m_{\phi}^2+\frac{k^2}{a^2}-\frac{2\phi}{M^2}V_{\rm eff,\phi}+\frac{2\dot{\phi}^2}{M^2}\right)\delta\phi=0\nonumber
\end{align}
In studying the perturbations it is convenient to introduce the
growth index
\begin{align}
\gamma(z,k) = \frac{\log\left(\frac{d\log\delta_m}{d\log
a}\right)}{\log(\Omega_m(z))}.
\end{align}
In $\Lambda$CDM we have $\gamma \approx 0.55$ (for $0.2\lesssim
\Omega_m \lesssim 0.3$), which is scale and almost redshift
independent. In modified theories however, $\gamma$ can have
significant scale and redshift dependence as shown in
\cite{chameleon_signatures,gamma_fofr,brax_shaw_fofr,brax_mod,mota_chameleoncosmology} for
the case of chameleon/$f(R)$ models.
\\\\
If we assume that the field is rolling slowly along the minimum we can neglect all terms proportional to $\dot{\phi}$ and the oscillating term $V_{\rm eff,\phi}$. The perturbations in $\phi$ will evolve more slowly than the perturbations in $\delta_m$ for scales deep inside the Hubble radius, thus, the term $\rho_m\beta,_{\phi}\delta_m$ and $(m_{\phi}^2+\frac{k^2}{a^2})\delta\phi$ will dominate over the $\delta\phi$ time derivatives in Eq.~(\ref{phi_eq}). Under these assumptions, we can simplify Eq.~(\ref{deltam_eq}) to
\begin{align}\label{quasistatic_approx}
\ddot{\delta_m} + 2H\dot{\delta_m} &= \frac{3}{2}\Omega_m H^2\frac{G_{\rm eff}}{G}\delta_m\\
\frac{G_{\rm eff}}{G} &= 1+\frac{2\beta^2\phi^2/\phi_0^2}{1+\frac{a^2}{\lambda_{\phi}^2k^2}}
\end{align}
which are the equations we use to integrate the perturbations.

At the time before SSB we have $\phi \approx 0$ and therefore $\frac{G_{\rm eff}}{G} = 1$. After SSB the field approaches the minimum $\phi_0=\pm\frac{\mu}{\sqrt{\lambda}}$, in this regime we have
\begin{align}
\frac{G_{\rm eff}}{G} = \left\{
\begin{array}{ll}
1 & \frac{a}{k} \gg \lambda_{\phi}\\
1+2\beta^2 &  \frac{a}{k} \ll \lambda_{\phi}
\end{array}
\right.
\end{align}
Thus small scales will feel a stronger gravitational constant.

\begin{figure*}
\centering
\includegraphics[width=0.9\columnwidth]{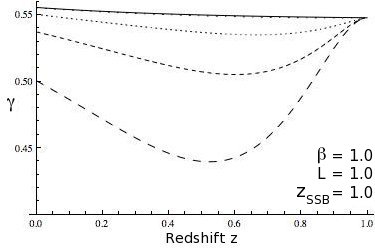}   \includegraphics[width=0.9\columnwidth]{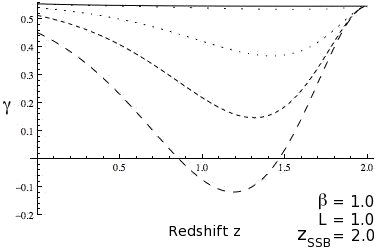}\\
\includegraphics[width=0.9\columnwidth]{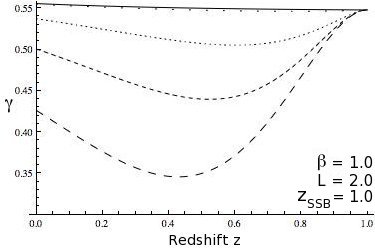}   \includegraphics[width=0.9\columnwidth]{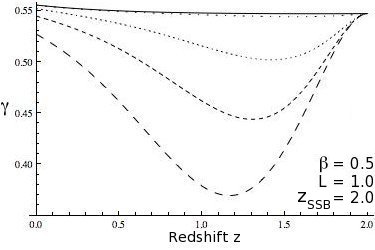}
\caption{The evolution of the growth index $\gamma(z)$ when $z_{\rm SSB} = 1$ (left) and $z_{\rm SSB}= 2$ (right) for four different wavenumbers $k = \{0.01,0.05,0.1,0.2\} \text{Mpc/h}$ (from top to bottom in each figure). The solid line show the predication of $\Lambda$CDM.}
\label{gamma_z}
\end{figure*}

\begin{figure*}
\centering
\includegraphics[width=0.9\columnwidth]{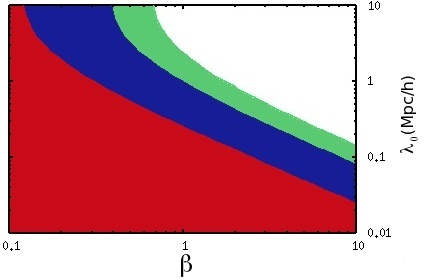}  \includegraphics[width=0.9\columnwidth]{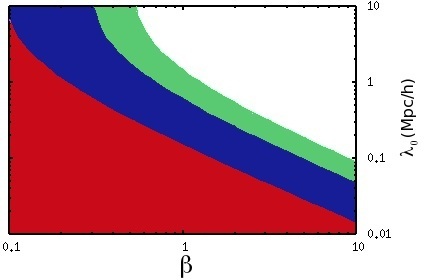}\\
\includegraphics[width=0.9\columnwidth]{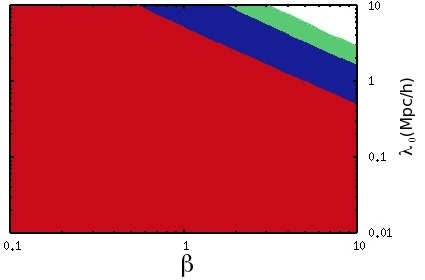}   \includegraphics[width=0.9\columnwidth]{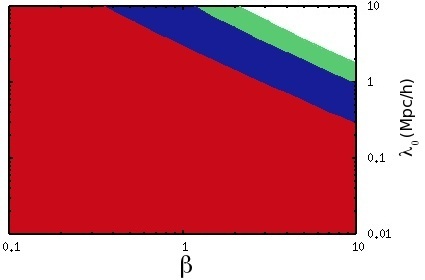}
\caption{The growth index $\gamma(z=0)$ for $z_{\rm SSB} = 1.0$ (left) and $z_{\rm SSB} = 2.0$ (right) for two comoving wavnumbers: $k = 0.2 \text{Mpc/h}$ (above) and  $k = 0.01 \text{Mpc/h}$ (below). The red region shows the GR regime $\gamma \simeq 0.555$, the blue region show the regime where $0.5 <\gamma < 0.55$, the green region shows $0.4 <\gamma < 0.5$ and the white region shows $\gamma < 0.4$.}
\label{gamma_z1_z2}
\end{figure*}

In Fig.~(\ref{gamma_z}) we show the redshift evolution of $\gamma$ for several different wavenumbers and in Fig.~(\ref{gamma_z1_z2}) we show contour plots for $\gamma(z=0)$ for two comoving wavenumbers.
\\\\
The growth rate on really large scales ($k \lesssim 0.01h/\text{Mpc}$) is not affected by the symmetron fifth force unless $L,\beta \gg 1$.
However on the smallest, linear scales we can still have a deviation from the predications of GR. Note that we have integrated the perturbations using the approximation Eq.~(\ref{quasistatic_approx}) instead of the full equations Eq.~(\ref{deltam_eq}-\ref{phi_eq}). The explanation for this is given in Sec.~(\ref{tac_ins}).

\subsection{Linear power spectrum and the CMB}

\begin{figure}
\centering
\centerline{\includegraphics[width=\columnwidth]{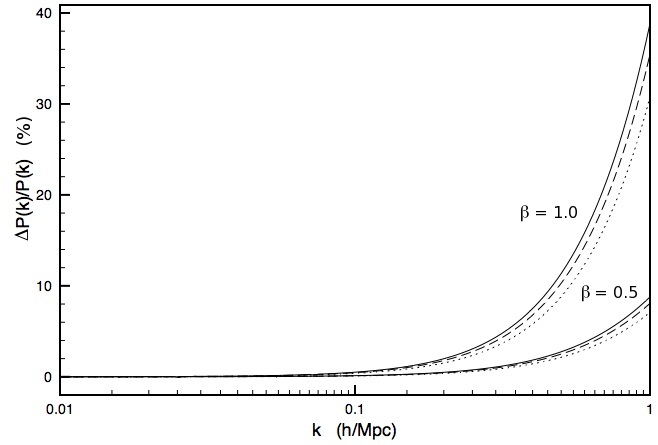}}
\caption{The linear power spectrum relative to that of $\Lambda$CDM for three different SSB redshifts: $z_{\rm SSB} = 0.5$ (dotted), $z_{\rm SSB} = 1.0$ (dashed) and $z_{\rm SSB} = 2.0$ (solid). We have fixed $L=1$ and shown the results for the two values $\beta = 0.5$ and $\beta = 1.0$.}
\label{linear_ps}
\end{figure}

In Fig.~(\ref{linear_ps}) we show the factional difference of the
linear power spectrum of the symmetron to that of $\Lambda$CDM,
defined as $\frac{\Delta P(k)}{P(k)} \equiv \frac{P(k)-P_{\Lambda
CDM}(k)}{P_{\Lambda CDM}(k)}$. Notice that on linear scales
$(k\lesssim 0.1 h/\text{Mpc}$) the power spectrum is very close to
$\Lambda$CDM. Going down to scales comparable to the length scale
of the symmetron ($k \sim \frac{1}{L} h/\text{Mpc}$) the
power spectrum starts to deviate significantly. However, in this
regime the perturbations are already nonlinear and we cannot
trust the results of the linear perturbation theory. Once we
discuss the N-body results we will see that the symmetron mechanism is
at work in this regime, thereby suppressing the predication of
linear perturbation theory.

The relative short range of the fifth force means that it will not
affect the CMB unless $L,\beta \gg 1$. Take $L=1$ and $\beta = 2$
as an example: we find a maximal increase in power (due to the ISW
effect) of $\sim 0.25 \%$ for multipoles around $l\sim 100$. One
needs a much larger $\beta$ and/or $L$ to have a detectable
signature in the CMB. The second case is not allowed by local
experiments while the first case implies a growth rate of the
linear perturbations which should have difficulty satisfying
constraints coming from large scale structure surveys.

A more thorough analysis of the linear perturbations in the symmetron model can be found in \cite{symmetron_brax}. There it was shown that strong signatures appears in other interesting linear observables such as the weak lensing slip parameter and the modified gravity parameter.

\subsection{Tachyonic Instability in the perturbations}\label{tac_ins}

The perturbations in Sec.~(\ref{linear_pert_section}) were
integrated using the approximate equation
Eq.~(\ref{quasistatic_approx}), which is equivalent to using the
analytical minimum as the background field. The reason we did not
use the full equations is because perturbations theory breaks down
close to $z_{\rm SSB}$.

Immediately before $z_{\rm SSB}$ the field is still close to $\phi = 0$;
as $z\to z_{\rm SSB}$ the mass of the field vanishes. This
means that the field cannot follow the minimum and starts to lag
behind as seen in Fig.~(\ref{phi_back}). The global minimum of the
effective potential $\phi = 0$ now becomes a local maximum and the
mass squared of the field becomes negative leading to an
exponential growth in the perturbations.

To see what happens we can simplify Eq.~(\ref{phi_eq}) by discarding
all but the most important terms.
\begin{align}\label{tachyonic_eq}
\ddot{\delta\phi} +  \left(m_{\phi}^2 + \frac{k^2}{a^2}\right)\delta\phi \simeq 0
\end{align}
If $m_{\phi}^2 + \frac{k^2}{a^2} < 0$ then the solution to the
above equation reads $\delta\phi \propto e^{t\sqrt{|m_{\phi}^2 +
k^2/a^2|}}$ which is exponentially growing.

In a realistic situation the field would roll very quickly down
from the false minimum $\phi = 0$, making $m_{\phi}^2$ positive,
and thereby stabilizing the field close to the symmetry breaking
minimum \citep{ssb}. Since perturbation theory is only valid
as long as the perturbations $\delta\phi$ are small we get a
breakdown of the perturbation theory when using the true
background solution. The blow up in $\delta\phi$, in turn, leads
to a blow up in the matter perturbations and the numerical results
cannot be trusted.

We have investigated this further by using N-body simulations. In
Fig.~(\ref{slice3}) we see a snapshot of the $\phi$-distribution
both before and after $z=z_{\rm SSB}=2.0$. There we see the same
sort of behavior as is familiar from symmetry breaking in
condensed matter physics: symmetry breaking takes place at
different places at different times according to the local matter
density. This type of dynamics is not taken care of in the
standard perturbation theory approach which leads to the apparent
instability.

Note that by using the analytical minimum when integrating the
perturbations we do not have control over the accuracy of our
results. A full analysis of this phenomenon could be handled with
N-body simulations, but in our simulations we have not explicitly
taken into account the time variation of the scalar field (we work
in the quasi-static limit) and our simulation box is also too
small to reach far enough into the linear regime. We leave this
study for future work.


\subsection{Varying Constants}

\subsubsection{WMAP constraints on particle mass variation}\label{BBN_bounds}

One important constraint on coupled scalar field theories comes
from time variation in the gravitational
constant $G$ in the Jordan frame, or equivalently in the masses of the standard model particles in the Einstein Frame. 
WMAP constrains any such variation to be less than about $5\%$ since
recombination \citep{bbnbounds}. Light element abundances provide similar constraints between the time of nucleosynthesis and today \citep{Accetta1990146}.
\\\\
Due to the conformal coupling to matter, $A(\phi)$, a constant mass scale in the Jordan frame becomes time and space varying in the Einstein frame. The mass variation between today and recombination is given by
\begin{align}
\frac{\Delta m}{m} = \frac{A(\phi_{\rm rec}) - A(\phi_{\rm today})}{A(\phi_{\rm rec})} \simeq \frac{1}{2}\left(\frac{\phi_{\rm today}}{M}\right)^2
\end{align}
where we have put $\phi_{\rm rec} \simeq 0$ since $z_{\rm rec} \gg z_{\rm SSB}$ in all interesting cases. If we further assume $\phi_{\rm today} = \phi_0$ we get the conservative constraint
\begin{align}
\frac{\Delta m}{m} \simeq \frac{1}{2}\left(\frac{\phi_0}{M}\right)^2 = \beta^2\left(\frac{M}{M_{\rm pl}}\right)^2 < 10^{-6}\beta^2
\end{align}
The WMAP constraint $\left|\frac{\Delta m}{m}\right| \lesssim 0.05$ is satisfied for all $\beta \lesssim 100$.

\subsubsection{Fine structure constant}

Analysis of absorption spectra of quasars have led some to claim that the fine structure constant $\alpha$ might have evolved by approximately one part in $10^5$ over the redshift range $0.2 < z < 3.7$. If this turns out to be true, then general covariance would imply that $\alpha$ can vary both in space and in time, that is, it must be a function of a field.

Since we have so far assumed that the symmetron couples conformally to matter fields, and since the Maxwell action is conformally invariant, at tree level the symmetron does not lead to a time varying $\alpha$. By considering a coupling of the symmetron to photons of the form
\begin{align}\label{photon_coupling}
S_{\gamma} = -\frac{1}{4}\int d^4x \sqrt{-g} A_{\gamma}(\phi) F_{\mu\nu}F^{\mu\nu}
\end{align}
where
\begin{align}
A_{\gamma}(\phi) = 1 + \frac{\zeta_{\gamma}}{2}\left(\frac{\phi}{M}\right)^2
\end{align}
then variations in $\phi$ will lead to variations in $\alpha$. Here $\zeta_{\gamma}$ is the symmetron-photon coupling relative to the symmetron-matter coupling. The variation in the fine structure constant between Earth (E) and another place (S) in the Universe is given by
\begin{align}
\left|\frac{\Delta \alpha}{\alpha}\right| = \frac{A_{\gamma}(\phi_E)-A_{\gamma}(\phi_S)}{A_{\gamma}(\phi_E)} \simeq  \frac{\zeta_{\gamma}}{2}\left(\frac{\phi_S}{M}\right)^2
\end{align}
If S is a very low density environment where $\phi_S \approx \phi_0$ then
\begin{align}
\left|\frac{\Delta \alpha}{\alpha}\right| \simeq \zeta_{\gamma}10^{-6}\beta^2
\end{align}
which for $\mathcal{O}(1) \lesssim \beta,\zeta_{\gamma}$ is close to the reported detection.

However, the local density in most Ly-$\alpha$ emitting systems is usually much larger than the cosmological background density today (see e.g. \cite{cosmological_chameleon} and references therein), which implies $\phi_S \ll \phi_0$ and the above estimate becomes even smaller.

To be able to account for the reported claims we need $z_{\rm SSB}$ to be well before the observed redshift of these systems and/or these systems to be located in voids to produce the desired $10^{-5}$ effect. This makes it possible that the symmetron is responsible for the claimed variations, but most likely it will require a fine tuning $\zeta_{\gamma} \gg 1$. A more detailed analysis, as done in \cite{baojiu_varying_alpha_nbody}, is required to see if this is the case.
This is beyond the scope of this paper. 


\subsection{N-body Equations}

To implement the general relativistic equations Eq.~(\ref{eom_phi},\ref{eeq},\ref{ttot},\ref{geo}) in N-body simulations, it suffices to work in the nonrelativistic limits, since the simulations only probe the weak gravity regime and small volumes compared with the cosmos. We write the perturbed metric in the (flat) conformal Newtonian gauge as
\begin{align}
ds^2 = -a^2(1+2\Xi)d\tau^2 + a^2(1-2\Psi)dx^{\mu}dx_{\mu}
\end{align}
where $\tau$ is the conformal time and $x^{\mu}$ is the comoving coordinate.
\\\\
The scalar field equation of motion in terms of the perturbed quantities becomes
\begin{align}
&-(1-2\Xi)\phi'' + \nabla_{{\bf x}}^2\phi - \phi' \left( 2H(1-2\Xi) - \Xi' - 3\Psi' \right)\nonumber\\
&= a^2\left( \phi\left(\frac{\rho_m}{M^2}-\mu^2\right) + \lambda\phi^3 \right)
\end{align}
Taking the quasi-static limit of this equation, in which we can neglect terms such as $\Xi'$, $\Psi'$ and $H\phi'$ since the time derivative of a quantity is much smaller than its spatial gradient, and removing the background part we obtain
\begin{align}\label{eom_nbody_phi}
\nabla_{{\bf x}}^2\phi &\approx \frac{a^2}{M_{\rm pl}^2}\left(\rho_m\phi - \overline{\rho}_m\overline{\phi})\phi\right)\nonumber \\
&+ \frac{a^2}{M_{\rm pl}^2}\left(\mu^2(\overline{\phi} - \phi) + \lambda(\phi^3-\overline{\phi}^3)\right)
\end{align}
where we have also used the approximation $A(\phi) \approx 1$ to
simplify the equation further.
\\\\
The $(0,0)$-component of the Ricci tensor and the trace of the
total energy-momentum tensor in the perturbed quantities becomes
\begin{align}
&a^2R_0^0 \approx -\nabla^2_{{\bf x}}\Xi + 3\left(\frac{a''}{a} - H^2\right)(1-2\Xi)\nonumber\\
& - 3\Psi'' -3H(\Xi' + \Psi')\\
&T \approx -A(\phi)\rho_m - 4V(\phi) + \frac{1}{a^2}(1-2\Psi)\phi'^2
\end{align}
The $(0,0)$-component of the Einstein equation with the background
part removed gives the nonrelativistic Poisson equation
\begin{align}\label{eom_nbody_poisson}
\nabla_{{\bf x}}^2\Phi \approx 4\pi G\left(\rho_m -\overline{\rho}_m\right)a^3
\end{align}
where we have neglected the contribution from the potential ($V(\phi)- V(\overline{\phi})$), put $A(\phi) \approx 1$ and taken $\Phi = a\Xi$ for convenience.
\\\\
The equation of motion for the N-body particles follows from the geodesic equation and reads
\begin{align}
\ddot{\bf x} + 2H\dot{\bf x} = -\frac{1}{a^3}\nabla_{\bf x}\Phi - \frac{1}{a^2}\frac{\phi}{M^2}\nabla_{\bf x}\phi - \frac{\phi\dot{\phi}}{M^2}\dot{\bf x}
\end{align}
By rewriting this equation in terms of the conjugate momentum to ${\bf x}$, ${\bf p} = a^2{\bf x}$, we have
\begin{align}\label{eom_nbody_particles}
&\frac{d{\bf x}}{dt} = \frac{{\bf p}}{a^2}\\
&\frac{d{\bf p}}{dt} = -\frac{1}{a}\nabla_{\bf x}\Phi - \frac{\phi}{M^2}\left(\nabla_{\bf x}\phi + \dot{\phi}{\bf p}\right)
\end{align}
The equations Eq.~(\ref{eom_nbody_phi},\ref{eom_nbody_poisson},\ref{eom_nbody_particles}) are all we need to put into the N-body simulation code in order to study structure formation in the nonlinear regime.


\section{N-body simulations}

Below we describe the algorithm and model specifications  of the
N-body simulations we have performed. We also give results from
tests of the code to show that the scalar field solver works
accurately.

\subsection{Outline}

For our simulations we have used a modified version of the
publicly available N-body code MLAPM \citep{mlapm}. The
modifications we have made follow the detailed prescription of
\cite{baojiu_coupled_sfc}, and here we only give a brief
description. The MLAPM code has two sets of meshes: the first
includes a series of increasingly refined regular meshes covering
the whole cubic simulation box, with respectively 4,8,16,...,$N_d$
cells on each side, where $N_d$ is the size of the domain grid,
which is the most refined of these regular meshes. This set of
meshes are needed to solve the Poisson equation using multigrid
method or fast Fourier transform (for the latter only the domain
grid is necessary). When the particle density in a cell exceeds a
pre-defined threshold, the cell is further refined into eight
equally sized cubic cells; the refinement is done on a
cell by cell basis and the resulting refinement could have
arbitrary shape which matches the true equal density contours of
the matter distribution. This second set of meshes are used to
solve the Poisson equation using the linear Gauss-Seidel
relaxation scheme.

The symmetron field is the most important ingredient in the model
studied here, and we have to solve for it to obtain detailed
information about the fifth force. In our N-body code, we have
added a new scalar field solver. It uses a nonlinear Gauss-Seidel
scheme for the relaxation iteration and the same criterion for
convergence as the default Poisson solver in MLAPM. But it uses
V-cycle instead of the self-adaptive scheme in arranging the
Gauss-Seidel iterations.

The modified Poisson equation is then solved using nonlinear
Gauss-Seidel relaxation on both the domain grid and the
refinements. With the gravitational potential $\Phi$ and the
scalar field $\phi$ at hand, we can evaluate the total force on
the particles and update their momenta/velocities which are used
to advance the particles in space.


\subsection{Simulation Details}
The physical parameters we use in the simulations are as follows: the present dark-energy fractional energy density $\Omega_{\Lambda} = 0.733$ and $\Omega_m = 0.267$, $H_0 = 71.9 \text{km}/s/\text{Mpc}$, $n_s = 0.963$ and $\sigma_8 = 0.801$. We use a simulation box with size $64 \text{Mpc}/h$, in which $h = H_0/(100 \text{km}/s/\text{Mpc})$. We simulate 9 different models, see Table.~(\ref{simdet}) for the symmetron parameter values.

\begin{table*}
\centering
\begin{tabular}{|l|c|c|c|c|c|c|c|c|c|}
\hline
  Model: & {\bf A} &  {\bf B} &  {\bf C} &  {\bf D} &  {\bf E} &  {\bf F} &  {\bf G} &  {\bf H} & {\bf  $\Lambda$CDM} \\
  \hline
   $z_{\rm SSB}$ 	& 0.5 & 0.5& 1.0& 1.0& 2.0& 2.0& 1.0& 1.0& 0.0  \\
  $\beta$ 		& 0.5 & 1.0& 0.5& 1.0& 0.5& 1.0& 0.5& 1.0& 0.0 \\
  $L$ 	 		& 1.0 & 1.0& 1.0& 1.0& 1.0& 1.0& 2.0& 2.0& 0.0  \\
  \hline
\end{tabular}
\caption{The symmetron parameters used in our simulations.}
 \label{simdet}
\end{table*}

These parameters are chosen so that they predict local fifth forces which are of the same order of magnitude as allowed by current experiments and observations and are such that we can see the effect of the different parameters. Note that the energy density in the symmetron is always much less than that of dark energy and therefore does not alter the background cosmology which in all runs will be that of $\Lambda$CDM.

In all those simulations, the particle number is $256^3$, so that the mass resolution is $1.114 \times 10^9 \text{Mpc}/h$. The domain grid is a $128\times 128\times 128$ cubic and the finest refined grids have 16384 cells on each side, corresponding to a force resolution of about $12 \text{kpc}/h$. The force resolution determines the smallest scale on which the numerical results are reliable.  Our simulations are purely N-body, which means that baryonic physics has not been included in the numerical code.


\subsection{Initial conditions}

Initial conditions for the simulation was generated using GRAFIC2 \citep{Bertschinger:2001ng,Prunet:2008fv} by using the parameters described above, but where we also included baryons with a density parameter $\Omega_b = 0.045$ (and a dark matter density $\Omega_m = 0.267 - \Omega_b$). We use the same initial conditions for all the simulations in order to see clearly the effect of the symmetron compared with $\Lambda$CDM.

This choice needs some justification. First of all, we start the simulation at $z=49$, a time in which the symmetron has no effect on the growth of the perturbations. This means that the only change the symmetron will have on the initial conditions is on the value of $\sigma_8$ today which is used to normalize the perturbations. Since the symmetron field has a rather short range compared to the linear regime we do not expect a large effect on $\sigma_8$ for the range $L \lesssim \mathcal{O}(1)$ we are considering.

To check this assumption we integrated the perturbations and calculated the value of $\sigma_8$ (by normalizing to the CMB) for our simulation models and found that the model with $L=1$ that is furthest away from $\Lambda$CDM, namely $F$ in which $z_{\rm SSB}=2.0$, $L=1$ and $\beta = 1$, only has $\sigma_8 \simeq 1.01\sigma_8^{\rm LCDM}$ justifying the use of $\Lambda$CDM initial conditions.

If one is to consider models in which $L$ is much larger than 1 then this becomes an issue that should be dealt with properly.


\subsection{Code tests}

Before we run simulations we have to make sure that the
scalar field solver, which is the main modification to the MLAPM
code, works accurately by performing code tests for situations
where the outcome is known from analytical solutions.

\begin{figure}
\centering
\includegraphics[width=\columnwidth]{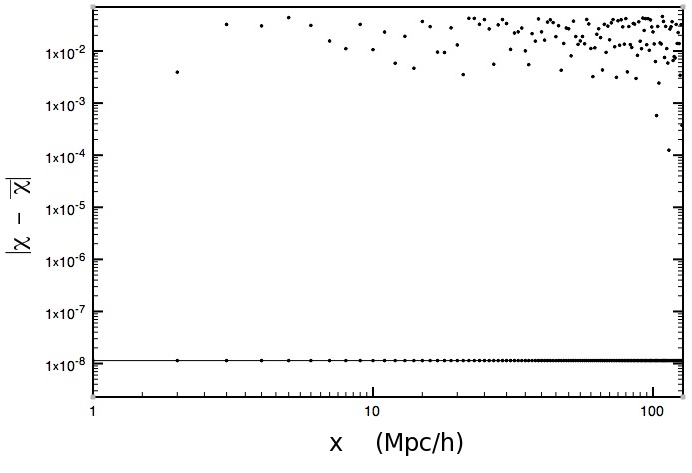}
\caption{The scalar field relative to the analytical solution before (random initial values above) and after (below) the Newton-Gauss-Seidel relaxation.}
\label{sf_test_homogenous}
\end{figure}

The scalar field solver uses the nonlinear Newton-Gauss-Seidel
relaxation scheme to compute $\chi \equiv \frac{\phi}{\phi_0}$,
and an indicator that it works is to show that, given the initial
guess of the solution that is very different from the true
solution, the relaxation could produce the latter within a
reasonable number of iterations. We consider a simulation box with
homogeneous density (obtained by putting particles on a regular
grid inside the simulation box), then the true solution is given
by $\chi = \overline{\chi}$: the background solution. We therefore
make an initial guess for $\chi$ which is randomly scattered
around $\overline{\chi}$ and let the scalar field solver solve for
$\chi$. The results for $|\chi - \overline{\chi}|$ before and
after the relaxation scheme are shown in
Fig.~(\ref{sf_test_homogenous}). The differences between the
initial guess and the true solution varies between 0.001 and 0.1
while after the relaxation the difference is of order $10^{-8}$.
By using double precision numbers in all the calculations we
obtained exactly the analytical solution (to double precision
$\approx 10^{-15}$), while for using only floating point numbers
the accuracy dropped to $10^{-6}$ which is exactly the accuracy in
floats. This shows that the scalar field solver works accurately.

\begin{figure}
\centering
\includegraphics[width=\columnwidth]{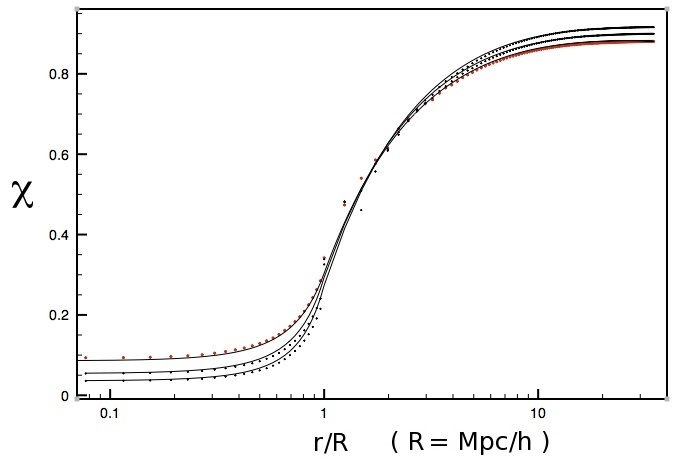}
\caption{The scalar field value as function as distance from the center for a spherical over-density embedded in a background of homogenous density $\rho_b$ together with the analytical solution for $\rho_c = 4000$, $6000$ and $8000$ times~$\rho_b$. The points shown here are calculated by binning the scalar field value using a bin-width $\Delta (r/R) = 0.01$ and taking the average. We used the same amount of particles, $128^3$, in each run so that the background density $\rho_b$ differs for the three cases shown above.}
\label{scalar_test_spherical}
\end{figure}

The most important effect of the symmetron is the screening
mechanism in which the local value of the field should be pushed
down towards $\chi = 0$ in high density environments. We therefore
consider a spherical over-density, located at the center of the
box, with a given radius $R$, homogeneous density $\rho_c$ inside
$R$ and embedded in a background of homogenous density $\rho_b$.
The analytical solution reads
\begin{align}
\chi(r) =& \chi(0) \frac{\sinh\left[m_c r\right]}{m_cr},~~~~~~~~~~~~~~~~~~~~~~~~r < R\\
\chi(r) =& \chi_b  + \frac{(\chi(R) - \chi_b)R}{r}e^{-m_b(r-R)},~~~~r> R
\end{align}
where
\begin{align}
&m_c^2  \simeq \left(\frac{\rho_c}{M^2}\right),~~~m_b^2  \simeq \left(\frac{\rho_b}{M^2}+ \mu^2(3\chi_b^2-1)\right)\nonumber\\
&\chi_b \simeq \sqrt{1 - \frac{\rho_b}{\rho_{\rm SSB}}},~~~\chi(R) =\chi_b\left(\frac{1+m_bR}{\frac{m_c R}{\sinh(m_c R)} + m_b R}\right)\nonumber\\
&\chi(0) = \chi_b\left(\frac{1+m_b R}{1+\frac{\sinh(m_cR)}{m_c R} m_bR}\right).
\end{align}
For the trial solution on the grid we use the background value
$\chi_b$ and we perform the test for a range of densities
$\rho_c$. The results after relaxation for the most massive cases
are shown in Fig.~(\ref{scalar_test_spherical}). There are some
small discrepancy from the analytical solution in the region $R <
r < 2R$ for the most extreme cases $\rho_c> 10^3 \rho_b$. This is
not a surprise as the density suddenly drops over 3 orders of
magnitude at $r=R$, meaning that we need a lot of particles in
this region in order to get accurate results. In the region $r<R$
and $r > 2R$ the scalar field solver produces the analytical
solution to high accuracy.



\section{Numerical Results}

In this section we present the results from the simulations, including the snapshots, the matter power spectrum and the halo mass function.


\subsection{Snapshots}
In the symmetron model $\chi = \frac{\phi}{\phi_0}$, and thereby the fifth force, is suppressed in high density regions. In this subsection we demonstrate these qualitative features using some snapshots.

\begin{figure*}[htbp]
\centering
\includegraphics[width=1\columnwidth]{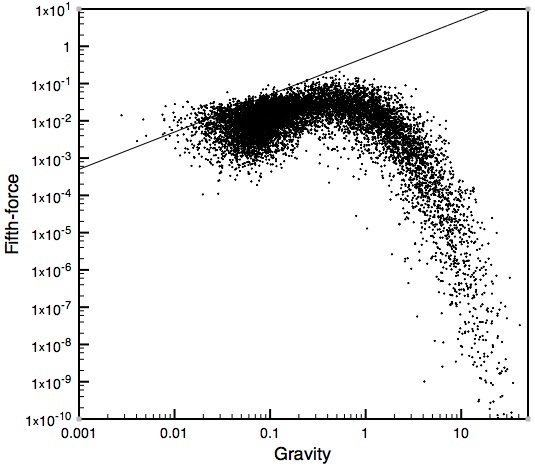}      \includegraphics[width=1\columnwidth]{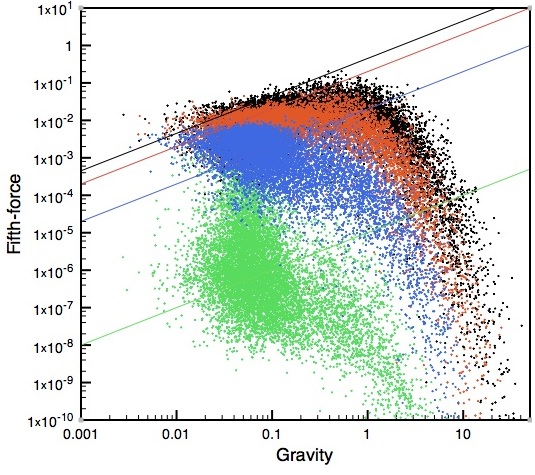}
\caption{The fifth force to gravity for in a slice of the simulation box at $z=0$ (left) and the comparison between four different redshift (right) for a runs with $z_{\rm SSB}=0.5$, $\beta=1$ and $L=1$. Black shows $z=0$, red $z=0.25$, blue $z=0.65$ and green $z=1$. The solid lines shows $\frac{F_{\phi}}{F_{\rm gravity}} = 2\beta^2\chi_b^2(z)$ which is the prediction for the (short-range) forces at the perturbation level (see Eq.~(\ref{quasistatic_approx})). Note that the force will be dispersed around this prediction because we already have significant over and under densities in which $\chi \not = \chi_b$. At $z=1$ the background field is close to $\chi_b = 0$ and the force is small everywhere in space. As we move closer to $z=0$, the symmetry breaks, and the background value moves towards $\chi_b = 1$. This means that the force in low density regions (small gravitational force) will increase whereas in high density regions (strong gravitational force) the screening kicks in and the force becomes suppressed just as seen above. The numerical size of the forces are given in terms of code units which are $\frac{H_0^2}{B}$ times the physical force unit.}
\label{forces}
\end{figure*}

Fig.~(\ref{forces}) shows the ratio of the fifth force to gravity
today for redshift both before and after $z_{\rm SSB}$.

At early times, the density is high everywhere and we expect the
fifth force on all particles to be strongly suppressed. At later
times we expect a screening in regions of high matter density.
These predictions are confirmed in Fig.~(\ref{forces}). We see
that fifth force on the particles which feel a strong
gravitational force (i.e. particles in a high density environment)
is highly suppressed whereas the fifth force on particles which
feel a weak gravitational force (i.e. particles in a low density
environment) follows the unscreened theoretical prediction
$F_{\phi} \simeq 2\beta^2\chi_b^2(z) F_{\rm gravity}$ (see
Eq.~(\ref{quasistatic_approx})).

\begin{figure*}
\centering
\includegraphics[width=1\columnwidth]{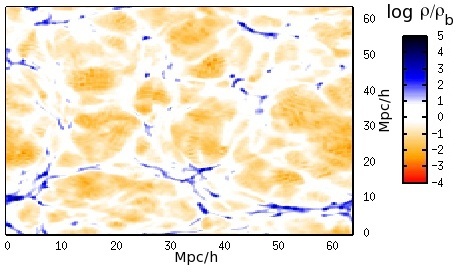}      \includegraphics[width=1\columnwidth]{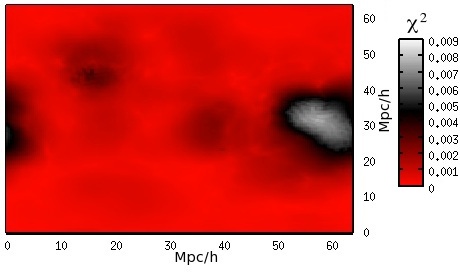}\\
\includegraphics[width=1\columnwidth]{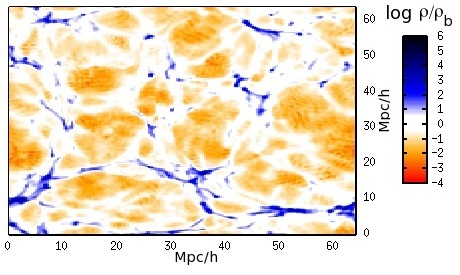}     \includegraphics[width=1\columnwidth]{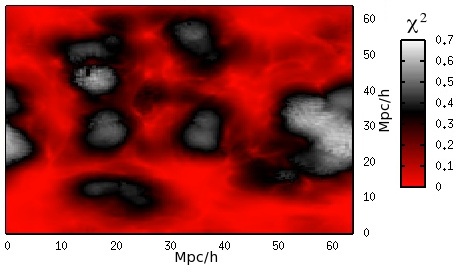}\\
\includegraphics[width=1\columnwidth]{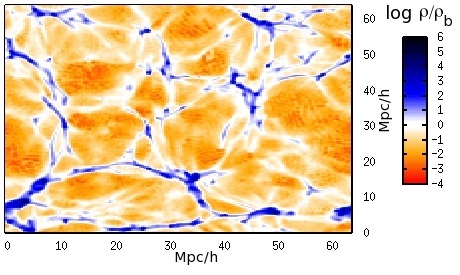}      \includegraphics[width=1\columnwidth]{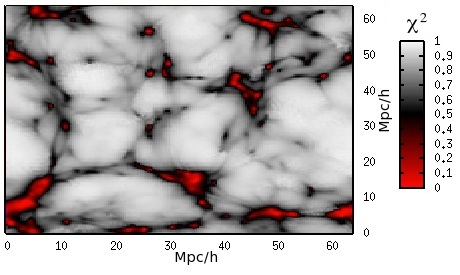}
\caption{The density distribution (left) and scalar-field distribution (right) for a run with $z_{\rm SSB} = 0.5$, $\beta=1$ and $L=1$. From top to bottom $z=1$,  $z=0.66$ and $z=0$. }
\label{slice1}
\end{figure*}

\begin{figure*}
\centering
\includegraphics[width=1\columnwidth]{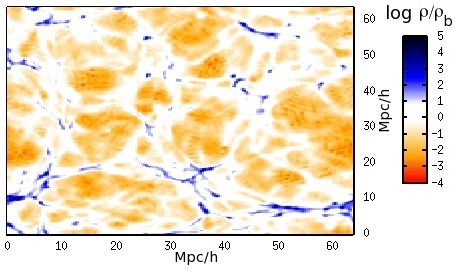}      \includegraphics[width=1\columnwidth]{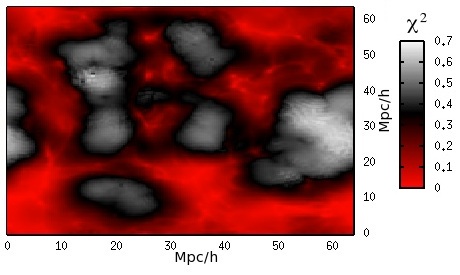}\\
\includegraphics[width=1\columnwidth]{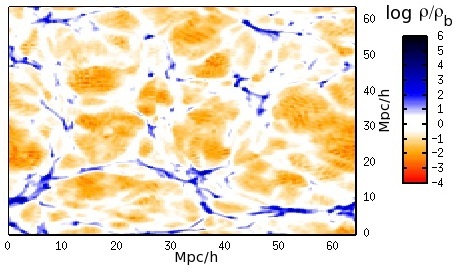}     \includegraphics[width=1\columnwidth]{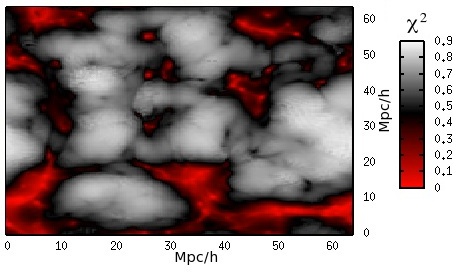}\\
\includegraphics[width=1\columnwidth]{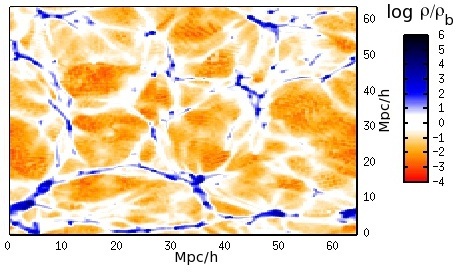}      \includegraphics[width=1\columnwidth]{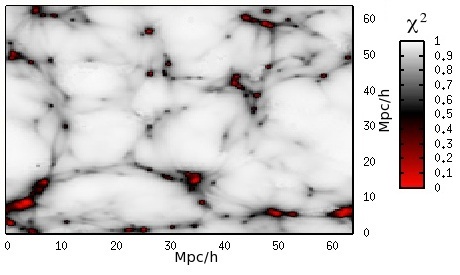}
\caption{The density distribution (left) and scalar-field distribution (right) for a run with $z_{\rm SSB} = 1$, $\beta=1$ and $L=1$. From top to bottom $z=1$,  $z=0.66$ and $z=0$.}
\label{slice2}
\end{figure*}

\begin{figure*}
\centering
\includegraphics[width=1\columnwidth]{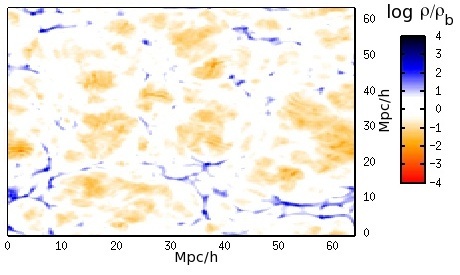}     \includegraphics[width=1\columnwidth]{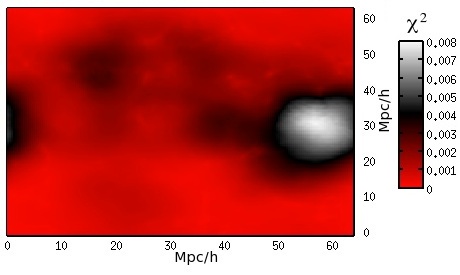}\\
\includegraphics[width=1\columnwidth]{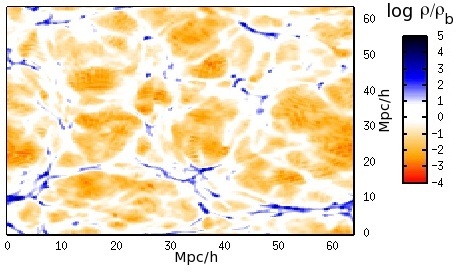}      \includegraphics[width=1\columnwidth]{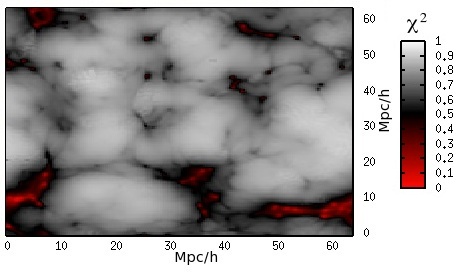}\\
\includegraphics[width=1\columnwidth]{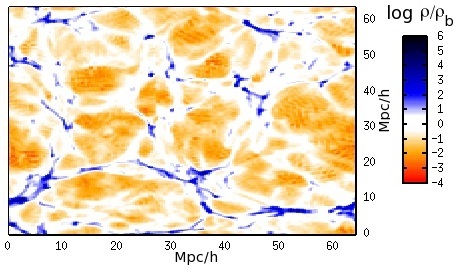}     \includegraphics[width=1\columnwidth]{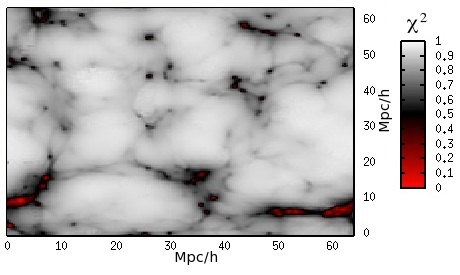}\\
\includegraphics[width=1\columnwidth]{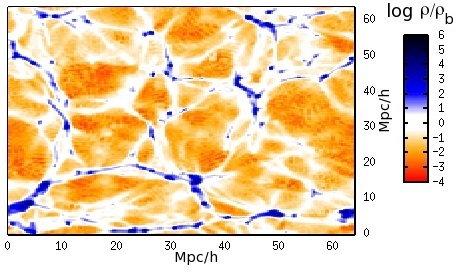}      \includegraphics[width=1\columnwidth]{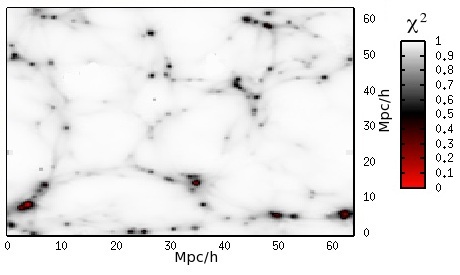}
\caption{The density distribution (left) and scalar-field distribution (right) for a run with $z_{\rm SSB} = 2$, $\beta=1$ and $L=1$. From top to bottom $z=2.33$, $z=1$,  $z=0.66$ and $z=0$.}
\label{slice3}
\end{figure*}

Fig.~(\ref{slice1})-(\ref{slice3}) shows the density and scalar
field distribution in a slice of the simulation box at different
redshifts for the three cases $z_{\rm SSB} =
0.5,1.0~\text{and}~2.0$ with $\beta = L = 1.0$ fixed.
\\\\
For redshifts $z > z_{\rm SSB}$, $\chi$ is very close to the
minimum $\chi = 0$ almost everywhere in space except in voids
where the symmetry has already been (weakly) broken. When we go
down to redshifts $z<z_{\rm SSB}$ the symmetry is broken in most
parts of the box, except in the high density regions where we still
have $\chi \sim 0$. Comparing the scalar field distribution today
for runs with different $z_{\rm SSB}$, we see that the earlier the
symmetry breaking takes place the larger the part of the box which
is unscreened ($\chi\sim 1$) today becomes. This is because the
critical density for the symmetry breaking is larger for larger
$z_{\rm SSB}$ and therefore the halos have to be more massive in
order to be effectively screened.



\subsection{Matter power spectrum}

The nonlinear matter power spectrum is an important observable and could be used to distinguish among different models of structure formation. As we have seen above the symmetron can have a strong effect on the growth rate of the linear perturbations for parameters that are allowed by local experiments. We expect these signatures to show up in the nonlinear matter power spectrum.

\begin{figure}[h]
\centering
\includegraphics[width=1\columnwidth]{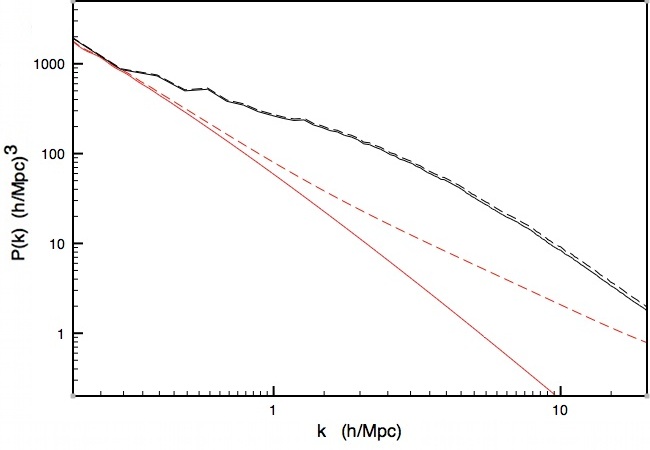}
\caption{The full nonlinear power spectrum for a run with $z_{\rm SSB} = L = \beta = 1.0$ (dashed black) and $\Lambda$CDM (solid black). For comparison we also show the corresponding predications from linear perturbations theory in red. We clearly see the effectiveness of the screening mechanism. The linear predications do not take the symmetron mechanism into account and are hugely overestimating the power on small scales relative to $\Lambda$CDM.\\\\}
\label{pow_all}
\end{figure}

\begin{figure*}
\centering
\includegraphics[width=1\columnwidth]{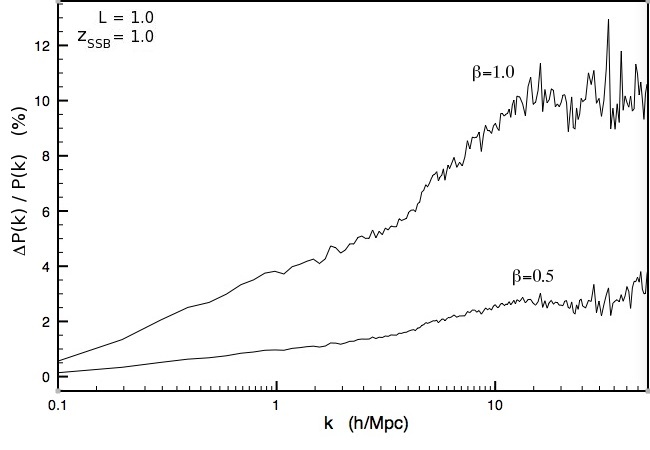}         \includegraphics[width=1\columnwidth]{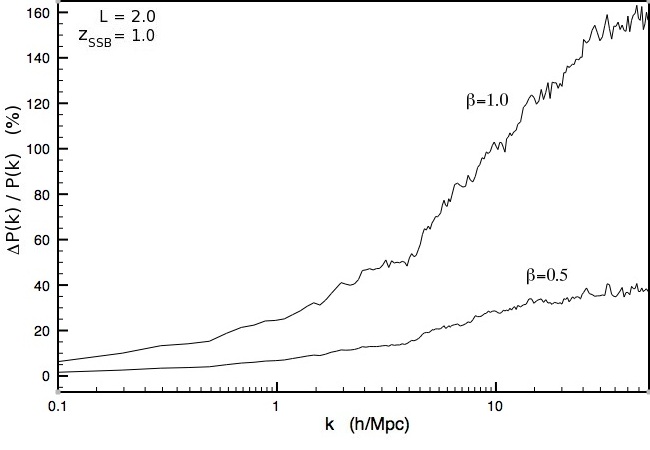}\\
\includegraphics[width=1\columnwidth]{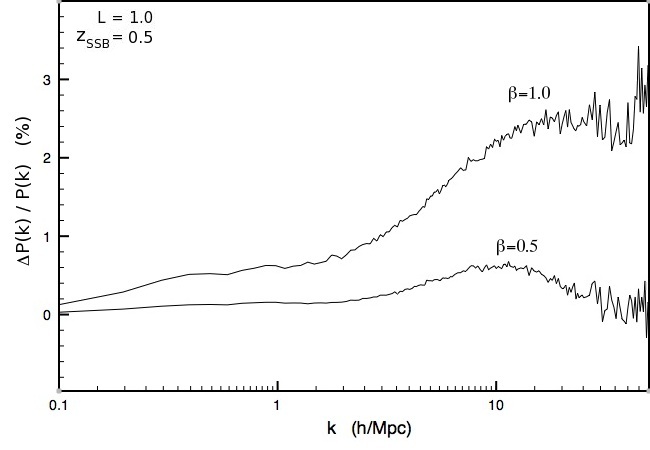}        \includegraphics[width=1\columnwidth]{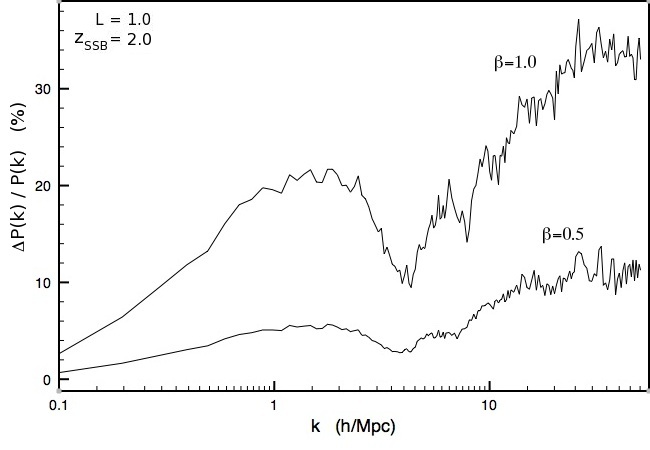}
\caption{The fractional difference in the nonlinear power spectrum relative to $\Lambda$CDM for $\{z_{\rm SSB} = 1.0, L = 1.0\}$  (top left), $\{z_{\rm SSB} = 1.0,L=2.0\}$ (top right), $\{z_{\rm SSB}=0.5,L=1.0\}$ (bottom left) and $\{z_{\rm SSB}=2.0,L=1.0\}$ (bottom right). For each case we show the results for the two values $\beta=0.5$ and $\beta=1.0$.}
\label{pow}
\end{figure*}

Fig.~(\ref{pow}) displays the fractional difference in the matter power spectrum from that of $\Lambda$CDM, defined as $(P(k)-P_{\Lambda CDM}(k))/P(k)$, and in Fig.~(\ref{pow_all}) we show the actual power spectrums for the symmetron and $\Lambda$CDM together with the corresponding predictions from linear perturbations theory.

The power spectrum agrees with the predictions of linear
perturbation theory on large scales ($k \lesssim
0.1~\text{Mpc}/h$), but on smaller scales the results found here
are weaker than the prediction of linear perturbation theory seen
in Fig.~(\ref{linear_ps}). This is because when linearizing the
field equation we are basically using the background matter
density everywhere and therefore preventing the symmetron
mechanism to take effect in suppressing the fifth force when
matter perturbations become large. In contrast, the N-body
simulation avoids this approximation by taking full account of the
suppression of the fifth force.

\begin{figure*}
\centering
\includegraphics[width=1\columnwidth]{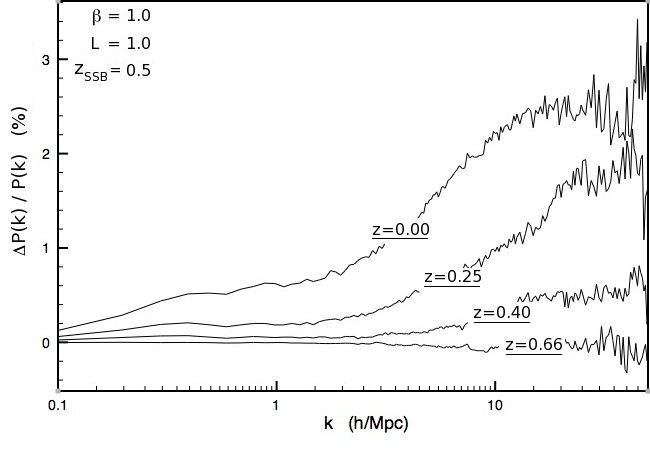}	\includegraphics[width=1\columnwidth]{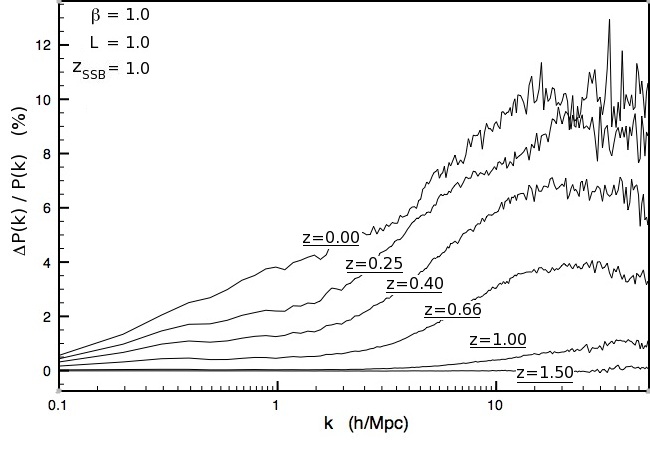}
\caption{The fractional difference in the nonlinear power spectrum relative to $\Lambda$CDM for $\{z_{\rm SSB} = 0.5, L = 1.0,\beta=1.0\}$  (left) and  $\{z_{\rm SSB} = 1.0,L=1.0,\beta=1.0\}$ (right) for several different redshifts.}
\label{pow_redshift}
\end{figure*}

The fractional difference relative to $\Lambda$CDM is growing with $z_{\rm SSB}$ and $\beta$ as the fifth force has more time to operate and is stronger. Comparing runs with the same $\beta$ we see an important effect if the symmetry 
breaking is earlier. When $z_{\rm SSB}=2.0$ the fractional power is increasing until we reach a scale where the screening mechanism becomes stronger and then starts to decrease again towards $\Lambda$CDM, only to start growing again at even smaller scales. This is because the critical density for having screening is much higher for larger $z_{\rm SSB}$ so that most halos (which are on small scales and of low mass) are unscreened.
\\\\
In Fig.~(\ref{pow_redshift}) we show the redshift evolution of the power spectrum. The power spectrum is found to be practically identical to that of $\Lambda$CDM for redshifts $z > z_{\rm SSB}$, but as soon as the symmetry breaks at the background level, the symmetron fifth force can kick in and enhance the clustering of matter.
\\\\
It is clear from Fig.~(\ref{pow}) that there exist a
large range of parameters in which the symmetron model can be
easily distinguished from $\Lambda$CDM.



\subsection{Halo profiles for $\chi$}

\begin{figure*}
\centering
\includegraphics[width=1\columnwidth]{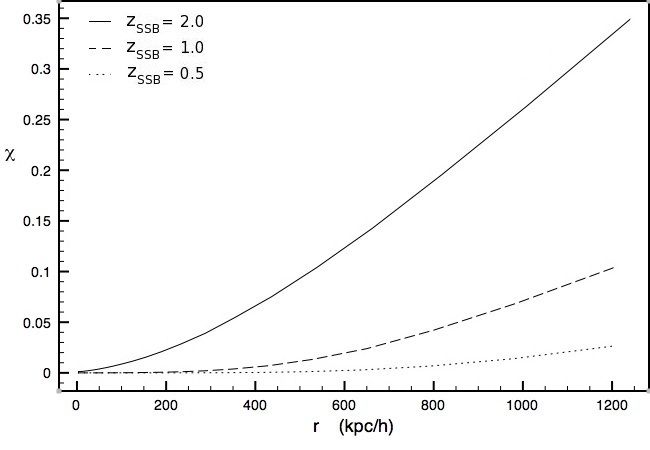}  \includegraphics[width=1\columnwidth]{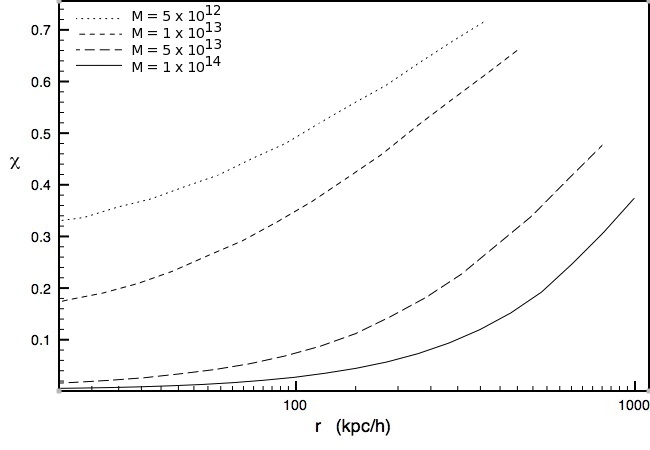}\\
\caption{Left: the halo profile of $\chi$ in the most massive halo of the simulation for three different symmetry breaking redshifts. Right: the halo profile of $\chi$ for four halos of mass (from top to bottom) $M=\{5\cdot10^{12}, 10^{13},5\cdot10^{13},10^{14}\} M_{\rm sun}/h$ in the same simulation where $z_{\rm SSB} = 2.0$. In both cases we have fixed $\beta = L = 1$.}
\label{haloprofile}
\end{figure*}

In Fig.~(\ref{haloprofile}) we show the profile of $\chi$ inside the most massive halos found in the simulation. Since the fifth force is proportional to $\chi$ this figure also provides information about the fifth force in halos.

The field profile of $\chi$ is seen to increase from the inner to
the outer regions of the halos and shows that the fifth force is
most suppressed in the central region as expected.

The fifth force is stronger for smaller halos, because those
generally reside in low density regions where the fifth force is
less suppressed. We see the closer the symmetry breaking redshift
is to $0$, the smaller $\chi$ becomes inside the halo and the more suppressed the
fifth force is. Again this is because early symmetry breaking means a
higher critical density and halo needs to be more massive
to be effectively screened. This effect is also seen in
Fig.~(\ref{slice1},\ref{slice2},\ref{slice3}) (notice the
difference in distribution of $\chi$ at $z=0$ between the
different runs) and also on the matter power spectrum in
Fig.~(\ref{pow}) (notice the way the power spectrum starts growing
again on small scales for $z_{\rm SSB} = 2.0$).

This has some important consequences for the local constraints. We mentioned in Sec.~(\ref{local_constraints}) that the local constraints were derived by assuming that our galactic neighborhood was not screened today, and lead to the constraint
\begin{align}\label{bound_lz}
L(1+z_{\rm SSB})^{3/2} \lesssim 2.3,
\end{align}
From our numerical results we see that when $z_{\rm SSB} = 2.0$ its only the most massive halos that are screened. This means that the assumption that went into the constraint above is very likely to be true. On the other hand for SSB that happens very close to today, halos of much smaller mass are in fact screened and it might be possible to have a range $L$ that exceeds Eq.~(\ref{bound_lz}) and still be in agreement with experiments.

We note that we have not seen any significant effect on the halo
density profiles. For a given mass range, the halo profiles seem
to have approximately the same distribution. There should be some
important differences for low mass halos, but the resolution in
our simulation is too low to study this.

However, the halo number counts was significantly different as we
shall see in the next section.

\subsection{Halo mass function}

The halo mass function $n$ is another key structure formation observable. It is defined to be the number density of dark matter halos within a given mass range. Because of the symmetron fifth force we expect more halos to be formed relative to the standard $\Lambda$CDM scenario.

We first look at the total number of halos (the integrated mass function) with more than 100 particles which clearly shows the effect of the fifth force, see Table.~(\ref{halonum}).

\begin{table*}
\centering
\begin{tabular}{|c|c|c|c|c|c|c|c|c|}
\hline
  {\bf A} &  {\bf B} &  {\bf C} &  {\bf D} &  {\bf E} &  {\bf F} &  {\bf G} &  {\bf H} & {\bf  $\Lambda$CDM} \\
  \hline
  1634 & 1694& 1678& 1871& 1758& 2051& 1671& 1788& 1607  \\  
  \hline
\end{tabular}
\caption{The halo count for our nine simulations. The corresponding symmetron parameters can be found in Table.~(\ref{simdet}).}
\label{halonum}
\end{table*}

\begin{figure*}
\centering
\includegraphics[width=1\columnwidth]{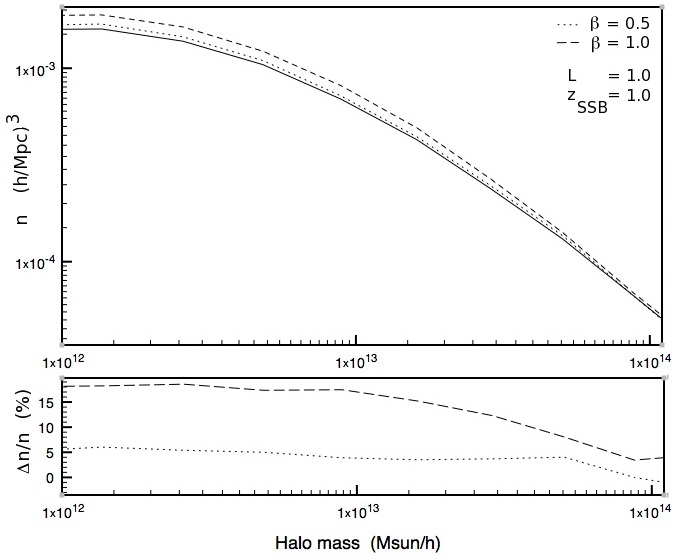}       \includegraphics[width=1\columnwidth]{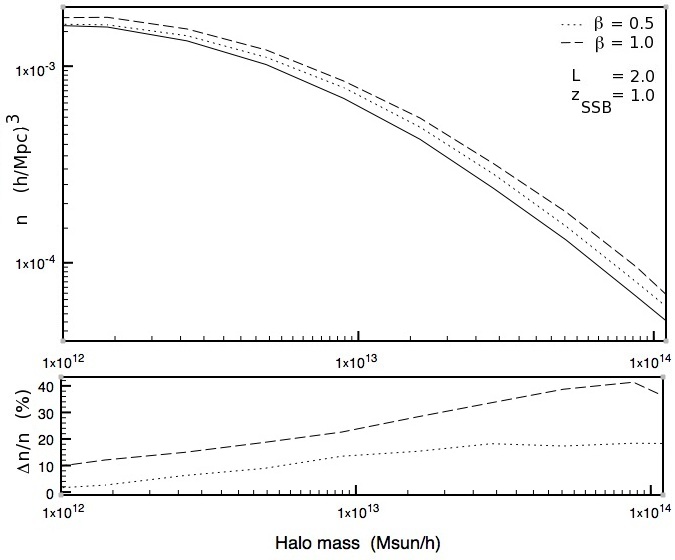}\\
\includegraphics[width=1\columnwidth]{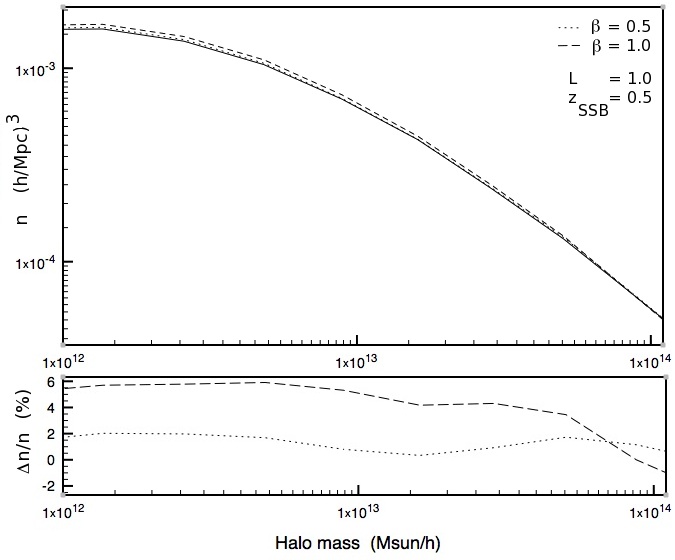}      \includegraphics[width=1\columnwidth]{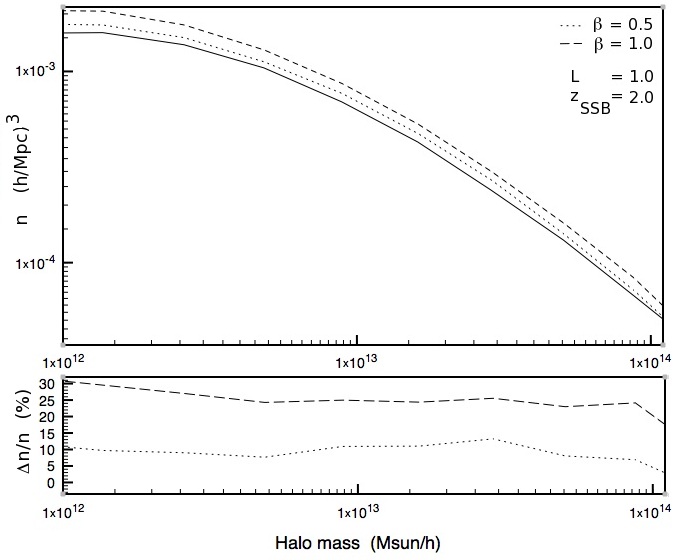}
\caption{The halo mass function for $\{z_{\rm SSB} = 1.0,L=1.0\}$ (top left),  $\{z_{\rm SSB} = 1.0,L=2.0\}$ (top right), $\{z_{\rm SSB} = 0.5,L=1.0\}$ (bottom left) and  $\{z_{\rm SSB} = 2.0,L=1.0\}$ (bottom right). The solid black line shows the prediction of $\Lambda$CDM  ($\beta=0$) and the dotted and dashes lines are for the two values $\beta=0.5$ and $\beta = 1.0$ respectively. We also show the fractional difference from $\Lambda$CDM. Note that we have smoothed the mass function over neighboring bins to remove noise arising from the binning as to show the trends more clearly.}
\label{halo}
\end{figure*}

In Fig.~(\ref{halo}) we have shown the mass function of the symmetron compared to $\Lambda$CDM at $z=0$. We see a significantly higher mass function, especially for low mass halos which are generally found in low density regions where the fifth force is unscreened. The earlier symmetry breaking occurs and the stronger the coupling strength $\beta$, the more halos are formed in agreement to what we would naively expect.

The mass function converges to that of $\Lambda$CDM at very large
halo masses for most parameters we have looked at. This is because
the most massive halos have taken a very long time to form and
therefore when the symmetron kicks in at some low redshift the
halo is already massive enough to be screened. However, for the
largest $z_{\rm SSB}=2.0$ we do have small increases in both the
halo number density and the mass of the most massive halos. There
have been reports of some tension between observations and
$\Lambda$CDM predictions with regards to very massive halos.
Unfortunately for the symmetron model to be able to elevate this
tension significantly we would need values of the parameters which
are in conflict with local experiments.

There is a large range of viable parameters for the symmetron where the mass function deviates significantly from $\Lambda$CDM.


\section{Summary and Conclusions}

The symmetron mechanism is one way a scalar field can be
non-minimally coupled to matter and still evade local gravity
experiments. The symmetron model suffers the same fine tuning
problems as chameleon models, but has the advantage of looking like
a more natural effective theory.

The energy density of the symmetron is too low to contribute to
the dark energy and we must therefore add a cosmological constant
to get accelerated expansion of the Universe. The background
evolution of the symmetron model is simply indistinguishable from
that of the $\Lambda$CDM model.

This degeneracy is broken by the linear perturbations. In
particular we have shown that the linear growth index
$\gamma(z,k)$ can have a significant scale and redshift dependence
together with a value today which can be distinguished from the
$\Lambda$CDM prediction for a large part of the parameter space.

The structure formation in the nonlinear regime was investigated
by using N-body simulations. N-body simulations have the advantage
over linear theory in its ability of fully capturing the
nonlinear environmental dependence of the symmetron field. Our
results confirm the expectation that in high density environments
the fifth force becomes screened. Consequently, the key
observables such as the nonlinear matter power spectrum is closer
to the $\Lambda$CDM predictions that expected from a linear
analysis.

We found that the symmetron can still produce large observable
signatures in both the nonlinear matter power spectrum and the
halo mass function, which could in principle be detected by
current and near future cosmological observations such as Euclid.

Note that in the simulations performed in this work, we have
treated baryons as dark matter. However, since the symmetron field
has a uniform coupling to all matter fields we expect that all the
results will qualitatively remain even after baryons are included.

In conclusion, the symmetron model has been found to have a wide
range of observable cosmological effects on both linear and
nonlinear scales. This adds to the list of observational
signatures like making galaxies brighter \citep{davis_galaxies}
and the possibility of being detected in near future local gravity
experiments \citep{khoury_symmetron} to mention some. The symmetron is therefore a
good candidate for the detection of new physics beyond the
standard model.


\section*{Acknowledgments}

The work described in this paper has been performed on TITAN, the
computing facilities at the University of Oslo in Norway. The
matter power spectrum was computed using POWMES
\citep{Colombi:2008dw} and the halo properties using MHF
\citep{Gill:2004km}. D.F.M. and H.A.W. thanks the Research Council
of Norway FRINAT grant 197251/V30. D.F.M. is also partially
supported by project PTDC/FIS/111725/2009 and CERN/FP/123618/2011.
H.A.W. thanks DAMPT at Cambridge University for the hospitality
where a part of this work was carried out. B.L. is supported by
Queens' College and the Department of Applied Maths and
Theoretical Physics of University of Cambridge. B.L. and A.C.D. thank STFC for
partial support. We would also like
to thank Douglas Shaw for useful discussions.


\section*{Appendix A: Useful Expressions}
Up to first order in the perturbed metric variables $\Xi$, $\Psi$ the non-zero components of the symmetric Levi-Civita connection are
\begin{align}
\Gamma^0_{00} &= \frac{a'}{a} + \Xi'\\
\Gamma^0_{0k} &= \Xi_{,k}\\
\Gamma^{i}_{00} &= \Xi^{,i}\\
\Gamma^{i}_{0k} &= \left(\frac{a'}{a}-\Psi'\right)\delta^{i}_{k}\\
\Gamma^0_{jk} &= \delta_{jk}\left(\frac{a'}{a}(1-2\Xi-2\Psi) -\Psi'\right)\\
\Gamma^{i}_{jk} &=-\Psi_{,k}\delta^{i}_{j} -\Psi_{,j}\delta^{i}_{k}+\Psi_,^{i}\delta_{jk}
\end{align}
From these expression we find that the components of the Ricci tensor and Ricci scalar are found to be
\begin{align}
R_{00} & = \Xi,^{i}_{i} - 3\left(\frac{a''}{a} - \left(\frac{a'}{a}\right)^2\right) + 3\Psi''\nonumber\\
& + 3\frac{a'}{a}(\Psi'+\Xi')\\
R_{0j} &= 2\Psi'_{,j} + 2\frac{a'}{a}\Xi_{,j}\\
R_{ij} &= -\Psi''\delta_{ij} -\frac{a'}{a}(\Xi'+5\Psi')\delta_{ij} - \Psi^{k}_{,k}\delta_{ij}\nonumber\\
&+ \left(\frac{a''}{a} + \left(\frac{a'}{a}\right)^2\right)(1-2\Psi - 2\Xi)\delta_{ij}\nonumber\\
&-(\Xi-\Psi)_{,ij}\\
R & = 6\frac{a''}{a^3}(1-2\Xi) + \frac{1}{a^2}(4\Psi^k_{,k} - \Xi^k_{,k})\nonumber\\
&-\frac{6}{a^2}\left(\Psi'' + \frac{a'}{a}(\Xi' + 3\Psi')\right)
\end{align}


\section*{Appendix B: Discretisation of Equations}
To implement the nonrelativistic equations into our numerical code, we have to rewrite then using code units, which are given by
\begin{align}
{\bf x_c} &= \frac{{\bf x}}{B},~~~{\bf p_c} = \frac{{\bf p}}{H_0B},~~~t_c = tH_0,~~~\chi = \frac{\phi}{\phi_0}\nonumber\\
\bf{\Phi_c} &= \frac{\bf{\Phi}}{(H_0B)^2},~~~\bf{\rho_c} = \frac{\rho_m}{\overline{\rho_m}},~~~\nabla = B\nabla_{\bf{x}}
\end{align}
where subscript $_c$ stands for code units, $B$ is the boxsize, $H_0=100\text{km}/s/\text{Mpc}$ and an overline denotes background quantities. In what follows we shall write $\nabla = \nabla_c$ for simplicity.
\subsection{Scalar field equation of motion}
The equation of motion for $\chi$ in code units becomes
\begin{align}
\frac{ac^2}{(BH_0)^2} \nabla^2 \chi  &\simeq a^3\left(\overline{\chi}-\chi + \chi^3 - \overline{\chi}^3\right)\left(\frac{\mu}{H_0}\right)^2\nonumber\\
&+ 3\Omega_m\left(\frac{M_{\rm  pl}}{M}\right)^2(\rho_c\chi - \overline{\chi})
\end{align}
where $\overline{\chi}$ is the background solutions and we have used $\phi_0^2 = \frac{\mu^2}{\lambda}$ to simplify. Note that $\chi$ varies in the region $0\leq \chi^2 \leq 1$. Discreticed this equation becomes $L^h(\chi_{i,j,k}) = 0$ where
\begin{align}
L^h(\chi_{i,j,k})   &= \frac{1}{h^2}\frac{ac^2}{(BH_0)^2}\left(\chi_{i+1,j,k}-2\chi_{i,j,k} + \chi_{i-1,j,k}\right)\nonumber\\
            & + \frac{1}{h^2}\frac{ac^2}{(BH_0)^2}\left(\chi_{i,j+1,k}-2\chi_{i,j,k} + \chi_{i,j-1,k}\right)\nonumber\\
            & + \frac{1}{h^2}\frac{ac^2}{(BH_0)^2}\left(\chi_{i,j,k+1}-2\chi_{i,j,k} + \chi_{i,j,k-1}\right)\nonumber\\
            & - a^3\left(\frac{\mu}{H_0}\right)^2\left(\overline{\chi} -\chi_{i,j,k}\right)\times\nonumber\\
            & \times\left(1- \chi_{i,j,k}^2 - \overline{\chi}\chi_{i,j,k} - \overline{\chi}^2 \right)\nonumber\\
            & - 3\Omega_m\left(\frac{M_{\rm pl}}{M}\right)^2\left(\rho_c\chi_{i,j,k} - \overline{\chi}\right)
\end{align}
The Newton-Gauss-Seidel iteration says that we can obtain a new and more accurate solution of $\chi_{i,j,k}^{\rm new}$ using our knowledge about the old solution $\chi_{i,j,k}^{\rm old}$ as
\begin{align}
 \chi_{i,j,k}^{\rm new} = \chi_{i,j,k}^{\rm old} -  \frac{L^h(\chi^{\rm old}_{i,j,k})}{\partial L^h(\chi^{\rm old}_{i,j,k}) / \partial \chi_{i,j,k}^{\rm old}}
\end{align}
where
\begin{align}
\frac{\partial L^h(\chi_{i,j,k})}{\partial \chi_{i,j,k}} &= - \frac{6}{h^2}\frac{ac^2}{(BH_0)^2}+ a^3\left(\frac{\mu}{H_0}\right)^2\left(1- 3\chi_{i,j,k}^2\right)\nonumber\\
&- 3\Omega_m\left(\frac{M_{\rm pl}}{M}\right)^2\rho_c
\end{align}
\subsection{Poisson Equation}
Since we can neglect the scalar field contribution to the Poisson equation, it remains unmodified from that of $\Lambda$CDM and reads (in code units)
\begin{align}
\nabla^2\Phi_c = \frac{3}{2}\Omega_m(\rho_{c,i,j,k}-1)
\end{align}
\subsection{Particle equation of motion}
Using the code units, Eq.~(\ref{eom_nbody_particles}) can be rewritten as
\begin{align}
\frac{d{\bf x_c}}{dt_c} &= \frac{{\bf p_c}}{a^2}\\
\frac{d{\bf p_c}}{dt_c} &= -\frac{1}{a}\nabla\Phi_c - \chi\left(\frac{\beta M}{M_{\rm pl}}\right)^2\left(\frac{c^2\nabla\chi}{(B H_0)^2} + \frac{d\chi}{dt_c}{\bf p_c}\right)
\end{align}
The factor $\left(\frac{M}{M_{\rm pl}}\right)^2$ can be also rewritten in terms of $L$, $\beta$ and $z_{\rm SSB}$ by using Eq.~(\ref{parameter_conversion}).\\\\


\bibliography{symmetron_nbody}
\end{document}